\documentclass[aps,prb,twocolumn,showpacs,10pt,superscriptaddress]{revtex4-1}

\usepackage{mathptmx}
\usepackage{graphicx}
\usepackage{subfigure}
\usepackage{amsmath}
\usepackage{amsfonts}
\usepackage{amssymb}
\usepackage{mathrsfs}
\usepackage{bbm}
\usepackage{subfigure}
\usepackage{xcolor}
\usepackage[colorlinks=true]{hyperref}
\usepackage{ulem}

\newlength{\seplinewidth}
\newlength{\seplinesep}
\colorlet{sepline}{orange}

\begin{document}

\title{Hysteresis from Nonlinear Dynamics of Majorana Modes in Topological Josephson Junctions}

\author{Jia-Jin Feng}
\altaffiliation{These authors contributed equally to this work.}
\affiliation{School of Physics, Sun Yat-sen University, Guangzhou 510275, China}

\author{Zhao Huang}
\altaffiliation{These authors contributed equally to this work.}
\affiliation{Texas Center for Superconductivity, University of Houston, Houston, Texas 77204, USA}

\author{Zhi Wang}
\email{wangzh356@mail.sysu.edu.cn}
\affiliation{School of Physics, Sun Yat-sen University, Guangzhou 510275, China}
\affiliation{Department of Physics, The University of Texas at Austin, Austin, Texas 78712, USA}

\author{Qian Niu}
\affiliation{Department of Physics, The University of Texas at Austin, Austin, Texas 78712, USA}

\begin{abstract}
We reveal that topological Josephson junctions provide a natural platform for the interplay between the Josephson effect and the Landau-Zener effect through a two-level system formed by coupled Majorana modes.
We build a quantum resistively shunted (RSJ) junction model by modifying the standard textbook RSJ model to take account the two-level system from the Majorana modes
at the junction. We show that the dynamics of the two-level system
is governed by a nonlinear Schr\"{o}dinger equation and solve the
equations analytically via a mapping to a classical dynamical problem. This nonlinear dynamics leads to hysteresis in the I-V characteristics, which can give a quantitative explanation to recent experiments.
We also predict coexistence of two interference patterns with periods $h/e$ and $h/2e$ in topological superconducting quantum interference devices.
\end{abstract}
\date{\today}
\maketitle


\section{Introduction}
The topologically protected degeneracy related to nonlocal nature of Majorana modes is among the core features of topological superconductors\cite{kitaev01,yakovenko,kitaevaip,zhangrmp}. This degeneracy is the foundation of fascinating topological qubits\cite{fu08,sato09,Tanaka09,sauprl10,alicea12,beenakker13,franzrmp,aliceaprx,aguadoreview} and also related to supersymmetry in condensed matter systems\cite{Qi09,Hsieh16,Huang17}. The situation is interesting as well when the degeneracy is split by couplings between Majorana modes\cite{beenakker08, Cheng09, Muzushima10,tewarijpcm,marcus16,prada17}. In particular for the one-dimensional case\cite{fuprb09,Lutchyn10,oregprl10,Mourik12,Deng12}, the split energy levels form a typical two-level system since other excitation levels have much higher energy\cite{alicea12,Platero12,aguado11}.

\begin{figure}[t]
\begin{center}
\includegraphics[clip = false, width =  \columnwidth ]{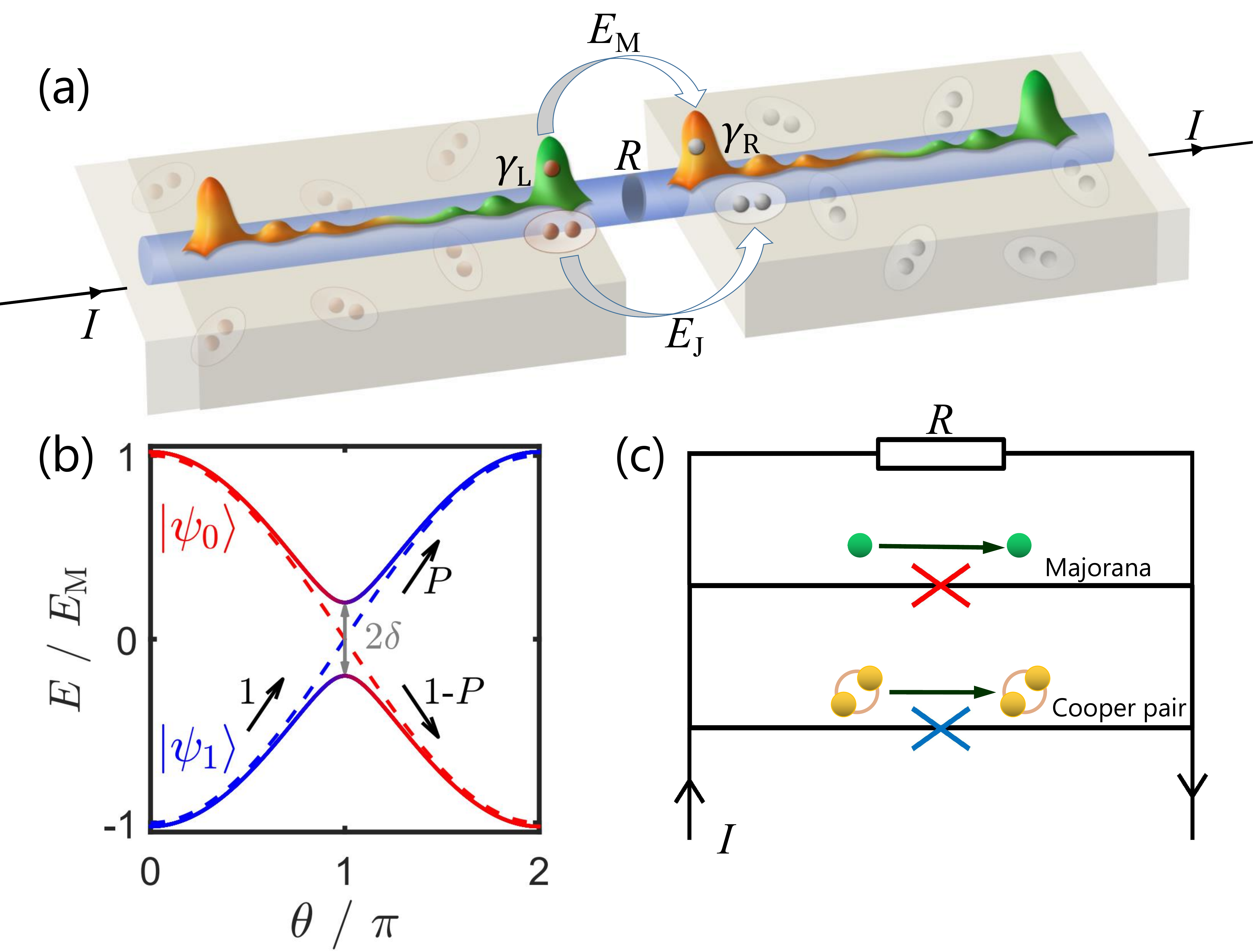}
\caption{(Color online) (a) Schematic of a topological Josephson junction with resistance $R$ driven by an injected current $I$. The single-electron tunneling through the Majorana modes $\gamma_L$ and $\gamma_R$, and the Cooper-pair tunneling induce Josephson couplings are quantified by energy scales of $E_{\rm M}$ and $E_{\rm J}$ respectively. (b) Energies of the two-level system defined by the two Majorana modes, with $\delta$ coming from the coupling between $\gamma_{L,R}$ and the other two Majorana modes at the ends of the wire. The Landau-Zener transition happens at the avoided energy crossing with $P$ the transition possibility. (c) Schematic of equivalent electric circuit for topological Josephson junction.}
\label{fig:setup}
\end{center}
\end{figure}

The two-level systems with their energy difference in control have proved extraordinarily fertile for interesting quantum phenomena\cite{AllenBook,Chuang05,Morsch06,noripr}. By coupling two Majorana modes with a Josephson junction as in Fig. \ref{fig:setup}a, two levels with energies $E\propto\pm\cos\theta/2$ are obtained, with $\theta$ the Josephson phase and the plus/minus signs correspond to states with opposite fermion number parity. Either level can coherently transport one electron through the junction, leading to the fractional Josephson effect $I \propto \pm \sin \theta/2$\cite{fuprb09,Lutchyn10,oregprl10}. In realistic systems where the two levels are inevitably coupled, the two-level system has avoided level crossings at $\theta = (2n+1)\pi$ as in Fig. \ref{fig:setup}b. Energy spectra with such avoided crossings are well known for the existence of the Landau-Zener (LZ) transitions\cite{LZ}: the two-level system enters a superposition state when the phase difference is driven by a finite voltage drop across the junction\cite{wangpra}. The topological Josephson junction thus hosts a natural platform for the interplay between the LZ effect and Josephson effect\cite{averin}. Since LZ effect has proved its impact on qualitatively changing the dynamics in various systems\cite{Chen11,Liu13,ludwig,higuchi,Law16,wubiao}, novel phenomena stemming from this interplay are expected on the topological junctions.

In this work, we study a realistic topological Josephson junction as sketched in Fig. \ref{fig:setup}a, where the supercurrent is contributed by tunneling in the form of both the single electron and Cooper pair. For a junction with negligible capacitance, we build a quantum resistively shunted junction (QRSJ) model by including the two quantum levels into the standard RSJ model. Under current injection, the two-level system can pass the avoided crossing again and again. At each passage it experiences a LZ transition at the near diabatic limit. 
The accumulation of multiple LZ transitions, which couple with the nonlinear dynamical of Josephson phase, induces a novel damped quantum oscillation.
We cast the quantum model into a classical model to solve this nontrivial dynamics by exploiting the method of averaging, and find that the LZ transitions are effectively described by a nonlinear Schr\"{o}dinger equation. We use phase-space portrait and the Poincar\'e map to analyze this nonlinear LZ effect, and reveal a separatrix which categorizes the dynamics into two distinct oscillatory behaviors. Within the separatrix, we obtain an analytically solution for the damped quantum oscillation, which agrees well with numerical simulations. We further show that this damped oscillation leads to hysteresis in the I-V curves, which gives a quantitative explanation to the recently reported ``unexpected'' hysteresis in HgTe topological Josephson junctions\cite{molenkamp,molenkamp2}.
We also predict, based on our theory, that in a topological superconducting quantum interference device (SQUID) two interference patterns with periods $h/e$ and $h/2e$ can coexist. This phenomenon will be an supporting evidence for Majorana modes if verified by future experiments.

\section{Quantum resistively shunted junction model}
The topological Josephson junction sketched in Fig. \ref{fig:setup}a consists of two topological superconductors, which could be one-dimensional nanowires with spin-orbit couplings\cite{marcus16}, superconducting quantum spin-Hall edge states\cite{kouwenhoven15}, or ferromagnetic atomic chains\cite{yazdani14}. The junction hosts two Majorana modes $\gamma_{\rm L, R}$ with their coupling described by\cite{kitaev01,fuprb09} $\mathcal{H}_{\rm M} = -i E_{\rm M} \gamma_{\rm L}\gamma_{\rm R}\cos({\theta}/{2})$ with $E_{\rm M}$ the maximum coupling energy. By defining a Dirac fermion $f = \gamma_{\rm L} + i \gamma_{\rm R}$, the Hamiltonian describes a typical two-level system where the empty state $|0\rangle$ and occupied state $|1\rangle$ are the two eigenstates. The corresponding energy spectra are $E_{\pm} = \pm E_{\rm M} \cos({\theta}/{2})$ which cross at $\theta = (2n+1)\pi$. In finite-size materials, the inevitable overlapping between $\gamma_{\rm L,R}$ and the other two edge Majorana modes leads to hybridization of the two states (see Appendix A for details), which produces avoided energy crossings. By writing the wave function as $|\psi \rangle = \psi_0  |0\rangle +  \psi_1  |1\rangle$, the dynamics is determined by the Schr\"{o}dinger equation
\begin{eqnarray}\label{eq:SE}
i \hbar \frac{\rm d}{{\rm d}t} \left(\begin{array}{cc}
\psi_0  \\
\psi_1
\end{array}\right) = \left(\begin{array}{cc}
E_{\rm M}\cos{\frac{\theta}{2}} & \delta\\
\delta & -E_{\rm M}\cos{\frac{\theta}{2}}
\end{array}\right) \left(\begin{array}{cc}
\psi_0  \\
\psi_1
\end{array}\right),
\end{eqnarray}
with $\delta$ the hybridization energy. This equation describes a two-level system which has an energy spectrum with avoided crossings at $\theta = (2n+1)\pi$,
as illustrated in Fig. \ref{fig:setup}b. When $\theta$ is driven through the avoided crossings, the LZ transition between the two levels will change the system from the ground state to the excited state with a textbook LZ transition probability $P= e^{ - 4\pi \delta^2/   (  \hbar \dot \theta E_{\rm M} ) }$.

We consider a junction with negligible capacitance, where the motion of $\theta$ under biased current can be described by the RSJ model\cite{tinkhambook,blackburn}, which is the current conservation equation where the total current $I$ is transported through the resistive and Josephson channel with $I= {V}/{ R } + I_{\rm J}$ as shown schematically in Fig. \ref{fig:setup}c. The Josephson current $I_{\rm J}$ has two parts: the conventional Cooper-pair channel $I_1 = I_{\rm c1}\sin{\theta}$, and the parity dependent Majorana channel $I_2  =  I_{\rm c2}\langle \psi| i\gamma_{\rm L}\gamma_{\rm R} |\psi\rangle \sin({\theta}/{2})$ which comes from the phase derivative of $\mathcal{H}_{\rm M}$\cite{fuprb09} (see Appendix A). By invoking the ac Josephson relation we obtain the equation explicitly as
\begin{equation}\label{eq:RSJ}
\frac{{\rm d}\theta}{{\rm d}t} = \frac{2eR}{\hbar} \left[I - I_{\rm c1}\sin{\theta} - I_{\rm c2}\left(|\psi_1|^2 - |\psi_0|^2\right)  \sin{\frac{\theta}{2}}\right],
\end{equation}
where the quantum average over Majorana operators is expressed with the wave function. This equation brings nonlinearity to the Schr\"{o}dinger equation (\ref{eq:SE}), and they together constitute the QRSJ model.

One important feature here is that when $I$ is large enough to make the right hand side of Eq. (\ref{eq:RSJ}) nonzero, the motion of $\theta$ would induce the LZ transitions around $\theta = (2n+1)\pi$. Different from the conventional LZ effect, the injected current drives the Josephson phase passing the avoided crossings again and again with a large velocity. Each time the LZ transition only induces a small change on the two-component wave function. However, the accumulation of many LZ transitions leads to a nonlinear quantum dynamics of the two-level system as we will show later. Therefore, the LZ transition is the building brick of the complicated but nontrivial dynamics of the two-level system in the topological junction.

\begin{figure}[b]
\begin{center}
\includegraphics[clip = true, width = \columnwidth]{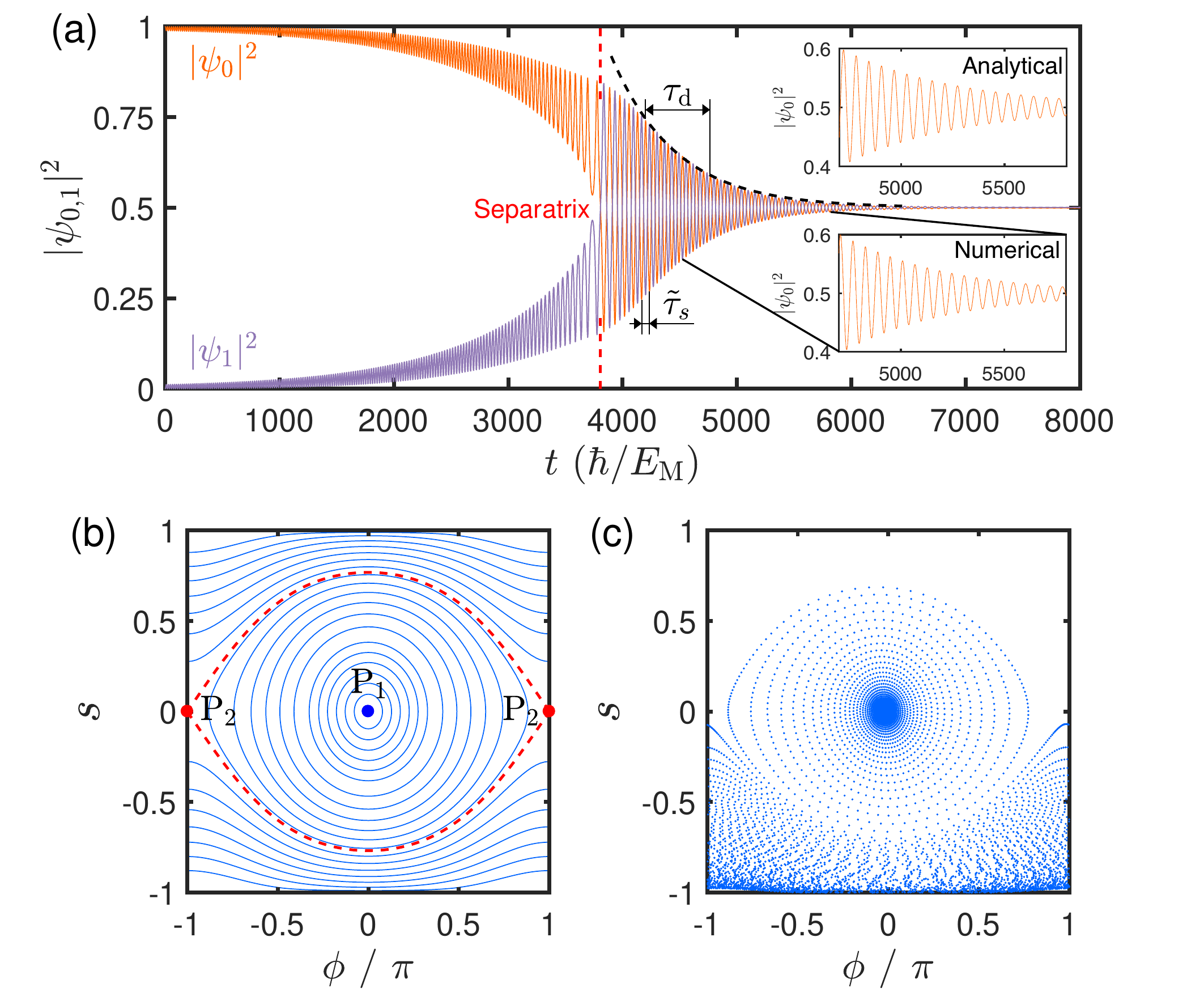}
\caption{(Color online) (a) Evolution of the wave function for the two-level system under constant injected current $I/I_{\rm c2}=1.5$, obtained by numerically solving Eqs. (\ref{eq:SE}) and (\ref{eq:RSJ}). The bottom inset is a zoom-in view in the marked time window. The analytical solution Eq. (\ref{eq:taudtaus}) provides the $\tilde \tau_s$, $\tau_d$, the dashed envelope line and the top inset. (b) Phase-space portrait of the classical Hamiltonian $H_c$, with $P_1$ the elliptic fixed point, $P_2$ the hyperbolic fixed point, and red-dashed circle the separatrix. (c) Poincar\'e map obtained by numerically solving Eq. (\ref{eq:classical}). Parameters of the junction are taken as $I_{\rm c1}/I_{\rm c2}=0.5$, $\delta/E_{\rm M}=0.02$, and $R=5\hbar/e^2$.}\label{fig:LZ}
\end{center}
\end{figure}

To observe the effect of these LZ transitions, we first numerically integrate Eqs. (\ref{eq:SE}) and (\ref{eq:RSJ}) with initial conditions $\psi_0=1$ and $\theta=0$, and present the time evolution of the wave function in Fig. \ref{fig:LZ}a.
We see that the wave function oscillates at the full time range. Looking carefully, the oscillation amplitude begins from a small value with the system mainly staying at $| 0 \rangle$, and then gradually increases. After passing a critical time marked by the red dashed line, the wave function begins to oscillate between $|0\rangle $ and $|1\rangle$. We will see later that this critical time relates to passing the separatrix of an effective classical Hamiltonian. We also notice that the oscillating period is shorter at the two ends of the time range, and becomes longer nearby the critical time.
Besides the rich oscillatory features, there is also an obvious damping on the envelope of the oscillations, with a characteristic time scale much larger than the oscillation periods. The damped quantum oscillation is unique and reflects the impact of the nonlinear dynamics of $\theta$ which enters the Schr\"{o}dinger equation of the two-level system.

\section{Nonlinear dynamics of Majorana Two-Level System}
Now we analyze this damped quantum oscillation by mapping the QRSJ model to a nonlinear classical model, which enables the usage of sophisticated approaches that have been developed for solving nonlinear classical dynamics\cite{dragoman,cao,Liujie,liujie2}. The trick is to notice that in the QRSJ model the wave function is subjected to two restrictions: it must be normalized, and the global phase is decoupled from the dynamics (see Appendix B for details). Then we can define two real variables: the relative amplitude $s = |\psi_1|^2 - |\psi_0|^2$ and the relative phase $\phi = {\rm arg}\psi_1 - {\rm arg}\psi_0$, which are complete for describing the dynamics of the two-level system\cite{Liujie,liujie2}. With this trick, we cast the QRSJ model into a purely classical model and write down the dynamical equations
\begin{subequations}\label{eq:classical}
\begin{align}
&\frac{{\rm d}\theta}{{\rm d}t} =  {2eRI} \left[1 - \frac {I_{\rm c1}}{I}\sin{\theta} - \frac{s I_{\rm c2}}{I}  \sin{\frac{\theta}{2}}\right] ,
\\
&\frac{{\rm d}s}{{\rm d}t} =  -{\delta} \sqrt{1-s^2} \sin \phi ,
\\
& \frac{{\rm d}\phi}{{\rm d}t} =  {E_{\rm M}} \cos \frac{\theta}{2} + \frac{ \delta s }{ \sqrt{1-s^2}}  \cos \phi,
\end{align}
\end{subequations}
where we take the unit $\hbar = 1$ for simplicity. Obviously the Eq. (\ref{eq:classical}a) is identical to Eq. (\ref{eq:RSJ}), and Eqs. (\ref{eq:classical}b) and (\ref{eq:classical}c) together are equivalent to Eq. (\ref{eq:SE}) which can be verified through simple algebra (see Appendix B for details). Here we have transformed the problem of  quantum dynamics to classical nonlinear dynamics in a three dimensional phase space.

With this mapping, the time scales of the system become clear as identified from the right hand side of Eq. (\ref{eq:classical}). We have $\tau_\theta = 1/2eRI,\ \tau_s = 1/\delta$ and $\tau_\phi = 1/E_{\rm M}$ which correspond to the change of $\theta,\ s$ and $\phi$. We note that these three time scales are different by orders with
$\tau_\theta \ll \tau_\phi \ll \tau_s$ for the junction parameters shown in Fig. \ref{fig:LZ}a and generally for $I>I_{\rm c1}+I_{\rm c2}$.

For classical nonlinear systems with multiple time scales, the method of averaging is a powerful technique\cite{smith}. The essence is to categorize "fast" variables and "slow" variables by typical time scales, then solve the equations for the fast variables by treating slow variables as constant parameters. After obtaining the solution, the fast variables are {\it averaged} over its time scale and used for solving the equations of the slow variables. With this process, the dynamical equations are decoupled into averaged equations, which significantly simplifies the problem.

Now we use the method of averaging to analyze the nonlinear dynamics in Eq. (\ref{eq:classical}), where $\theta$ is treated as the fast variable and $s, \phi$ as slow variables, since $\tau_\theta$ is the smallest time scale.
We first consider $s$ unchanged in $\tau_\theta$ and solve Eq. (\ref{eq:classical}a) to obtain the time average of $\cos \frac{\theta} {2}$, defined as $\overline \cos \frac{\theta} {2} \equiv   \int{\rm d}t\cos \frac{\theta} {2}$ with integration range the time for $\theta$ to rotate $4\pi$.

By taking the time derivative on both sides of Eq. (\ref{eq:classical}a), we can obtain terms containing $s$ and $\dot s$. Within $\tau_\theta$, because $s$ and $\dot s$ both vary slowly, we take them as time independent. With some tedious but straightforward computation we obtain (see Appendix C for details)
\begin{eqnarray}\label{eq:cosaverage}
\overline \cos ({\theta}/ {2})  \approx \alpha s +\beta \dot s ,
\end{eqnarray}
with $\alpha=I_{\rm c1}I_{\rm c2}/I^{2}$ and $\beta =  {I_{\rm c2}\tau_\theta}/{I}$ from the lowest order Taylor expansion of $I_{\rm c1}/I$ and $I_{\rm c2}/I$. Here $\alpha s$ is much larger than $\beta \dot s$, and we refer them as zeroth-order and first-order averaging respectively.

We begin from the zeroth-order averaging and replace $ \cos \frac{\theta} {2}$ with $\overline \cos \frac{\theta} {2}  = \alpha s$ in the Schr\"{o}dinger equation (\ref{eq:SE}), and obtain
\begin{small}
\begin{eqnarray}
i \hbar \frac{\rm d}{{\rm d}t}\!\! \left[\begin{array}{cc}
\!\!\psi_0 \!\! \\
\!\!\psi_1\!\!
\end{array}\right] \!\!=\!\! \left[\begin{array}{cc}
{\!\!E_{\rm M} \alpha (|\psi_1|^2 -|\psi_0|^2)} & \delta\\
\delta & \!\!\!\!\!\!\!\!\!\!\!\!\!\!-{E_{\rm M} \alpha (|\psi_1|^2 -|\psi_0|^2)\!\!}
\end{array}\right] \!\!\!\!\left[\begin{array}{cc}
\!\!\psi_0 \!\! \\
\!\!\psi_1\!\!
\end{array}\right],\label{eq:nonlinearSE}
\end{eqnarray}
\end{small}which becomes a typical nonlinear Schr\"{o}dinger equation due to the nontrivial diagonal elements \cite{Liujie,liujie2,Pitaevskii}. This explicitly shows that coupling to Josephson phase dynamics brings the nonlinearity into the quantum dynamics of the two-level system, which is the reason for the rich and unusual dynamical behaviors shown in Fig. \ref{fig:LZ}a (see Appendix C for details).

Now we interpret this nonlinear quantum dynamics with the classical model. In Eq. (\ref{eq:classical}c) by replacing $\cos\frac{\theta}{2}$ with its average , we obtain,
\begin{eqnarray}\label{eq:thetaout}
 \frac{{\rm d}\phi}{{\rm d}t}=  E_{\rm M}\alpha  s+ \frac{\delta s  }{ \sqrt{1-s^2}}  \cos \phi .
\end{eqnarray}
Now the system is only described by Eq.~(\ref{eq:thetaout}) and Eq.~(\ref{eq:classical}b) with $\theta$ integrated out. These two equations are the canonical equations of a classical Hamiltonian (see Appendix C for details),
\begin{eqnarray}\label{eq:classicalH}
H_{\rm c} =   - \frac{ 1}{ 2 } \alpha E_{\rm M}  s^2  + \delta \sqrt{1-s^2} \cos \phi,
\end{eqnarray}
where $s$ and $\phi$ are the coordinate and canonical momentum.

Let us use the phase space portraits of this effective Hamiltonian, as shown in Fig. \ref{fig:LZ}b, to understand the oscillatory features shown in Fig. \ref{fig:LZ}a. There is an elliptic fixed point $P_1$ at $(s,\phi)=(0,0)$, and a hyperbolic fixed point $P_2$ at $(s,\phi)=(0,\pm \pi)$ (see Appendix C for details). A separatrix connects the hyperbolic fixed point, separating the phase space into two distinct areas: extended trajectories outside the separatrix and orbiting trajectories around the elliptic fixed point inside the separatrix.

The extended trajectories outside the separatrix in Fig. \ref{fig:LZ}b correspond to dynamics before the critical time in Fig. \ref{fig:LZ}a. For motion along these trajectories, the $s$ stays negative or positive, agreeing with the small oscillations with $|\psi_0| > |\psi_1|$ at the beginning of Fig. \ref{fig:LZ}a.
Inside the separatrix, the trajectories become orbital, with $s$ oscillating from negative to positive values. This corresponds to the oscillations in Fig. \ref{fig:LZ}a after the critical time, where $|\psi_0|$ and $|\psi_1|$ have overlapped oscillations.
When approaching the separatrix, the period of the orbits is enlarged since the period should be divergent at the separatrix\cite{dittrich}. This corresponds to the observed period enlargement near the critical time in Fig. \ref{fig:LZ}a.
From the above analysis, we argue that the system begins from outside of the separatrix, passing through the separatrix at the critical time, and then orbits inside the separatrix and finally reaches the elliptic fixed point.

For clarity we demonstrate the Poincar\'e map of the numerical results for Eq. (\ref{eq:classical}) in Fig. \ref{fig:LZ}c, which is obtained by recording the points on the $s-\phi$ plane with $\theta = 4n \pi$. The local trace of the Poincar\'e map follows
the trajectories of the classical Hamiltonian, illustrating that the oscillations shown in Fig. \ref{fig:LZ}a can be approximately determined by the classical Hamiltonian. The global structure of the Poincar\'e map, however, demonstrates a spiral-in feature from outside the separatrix to the elliptic fixed point $P_1$. This exhibits the effect of a friction force which brings all phase-space trajectories to elliptic fixed points. This long-time-scale damping, also shown in Fig. \ref{fig:LZ}a, cannot be obtained based on the zeroth-order averaging.

Now we explore the damping feature by including the first-order averaging, and replacing $\cos\theta/2$ with Eq. (\ref{eq:cosaverage}). Around the elliptic fixed point $P_1$, we find that Eqs. (\ref{eq:classical}b) and (\ref{eq:classical}c) lead to  (see Appendix C for details)
\begin{eqnarray}\label{eq:harmonics}
\ddot s + {\beta E_{\rm M} } \dot s+  (\delta^2 + \alpha E_{\rm M } \delta) s  = 0,
\end{eqnarray}
which is nothing but a classical damped harmonic oscillator. It has a standard solution of the form,
\begin{eqnarray}\label{eq:analyticalsolution}
s & = & e^{-t/ {\tau_{\rm d}}}  \cos (2\pi {t}/{\tilde\tau_s}),
\end{eqnarray}
with the damping and oscillating time of
\begin{eqnarray}\label{eq:taudtaus}
\tau_{\rm d} =  \frac{2eRI^2 } { I_{\rm c2} E_{\rm M} \delta  }, \quad \tilde \tau_s = \frac{ 2\pi }{\delta \sqrt{ 1  + \alpha {E_{\rm M}}/{ \delta}}}.
\end{eqnarray}
We plot this analytical solution as an inset of Fig. \ref{fig:LZ}a, and find that it agrees well with the numerical simulations around the elliptic fixed point. Here we have demonstrated a duality between the nonlinear quantum dynamics in this two-level system and a classical damped harmonic oscillator which is exactly solvable. Therefore, this duality enables us to find an analytical solution for the damped quantum oscillations despite the equations for the nonlinear quantum dynamics is rather complicated. In fact, we further show a mapping to a solvable anharmonic damped oscillator (see Appendix C for details), which even correctly describes the dynamics far from the elliptic fixed point.

\begin{figure}[t]
\begin{center}
\includegraphics[clip = true, width =  \columnwidth]{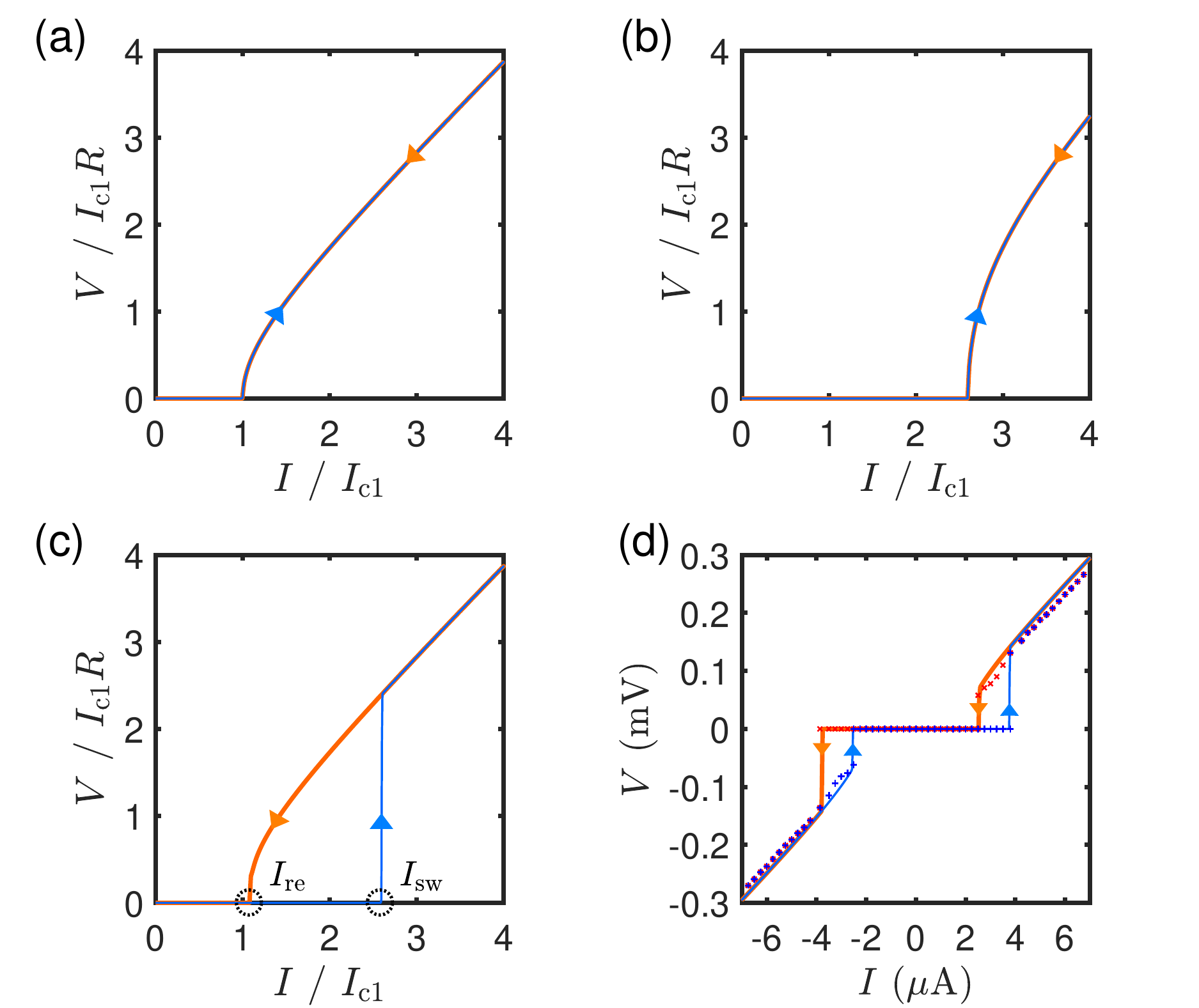}
\caption{(Color online) I-V curves in absence of LZ effect for (a) $I_{\rm c2} = 0$, (b) $I_{\rm c2}/I_{\rm c1} = 2$ and $\delta=0$. (c) I-V curves in presence of LZ effect with parameters the same as in Fig. \ref{fig:LZ}a.  (d) Comparison between our numerical simulation (solid lines) and the experimental data (discrete crosses) taken from Ref. [\onlinecite{molenkamp}]. Junction parameters in simulation are adopted the same as in the experiments with the resistance $R=44\Omega$, the capacitance $C=34$aF, the $2\pi$-period current $I_{\rm c1}=2\mu$A, the $4\pi$-period current $I_{\rm c2}=2.3\mu$A, and the decoherence time chosen as $\tau_2 = 10^5\hbar/E_{\rm M}$.}
\label{fig:qrcsj}
\end{center}
\end{figure}

\section{Hysteresis in I-V curves}
Now we study the I-V characteristics of the topological Josephson junction based on the QRSJ model. We numerically simulate the average voltage upon adiabatic current injection, which gradually increases to a large value and then decreases back to zero.
As a benchmark, we first show the I-V curve for a trivial junction with $I_{\rm c2} = 0$ in Fig. \ref{fig:qrcsj}a, which is the well known result of $V= R \sqrt{ I^2 - I^2_{\rm c1}}$ around the critical current \cite{tinkhambook}. We then consider an additional $4\pi$-period Josephson current $I_2=I_{\rm c2}\sin \frac{\theta}{2}$ which corresponds to the case of local parity conservation with $\delta = 0$, where LZ effect cannot take place. We solve the Eq.~(\ref{eq:RSJ}) with $|\psi_0|^2=1$ or $|\psi_1|^2=1$, and obtain the I-V curve as shown in Fig.~\ref{fig:qrcsj}b. Clearly the simple addition of a $4\pi$-period Josephson current modifies the shape of the I-V curve but demonstrates no novel phenomenon. For both cases, the voltage which is the velocity of the phase difference is fully determined by the applied current, so the quantum dynamics is history independent.

However, when $\delta$ becomes finite and the LZ transitions begin to affect the tunneling current, we find an unambiguous hysteretic I-V curve with two critical currents as shown in Fig. \ref{fig:qrcsj}c: a switching current $I_{\rm sw}$ where the voltage jumps from zero to finite value and a smaller retrapping current $I_{\rm re}$ for the finite voltage jumping back zero.

The origin of this hysteresis can be understood with the time evolution of $|\psi_0|^2$ and $|\psi_1|^2$ discussed in Fig. \ref{fig:LZ}a. Initially for a small injected current below the switch value, the voltage is zero and the two-level system stays at $|1\rangle$ with certainty ($|\psi_1|^2=1$). When the injected current is increased above the switching current, the probabilities begin to oscillate due to the nonlinear dynamics of the two-level system as detailedly discussed in previous section. The oscillation is strongly damped and after a while, the two-level system enters a state with nearly equal probability of the two levels since they are symmetric with the phase translation.

From above, we can see that the Josephson current is contributed by only one level for the zero-voltage stage but both levels for the finite-voltage stage. Therefore, it is reasonable that the critical currents are different when the injected current is increasing or decreasing. Because only one level contributes to the Josephson current as in the current increasing stage, we have $I_{\rm J}=I_{\rm c1}\sin{\theta}+I_{\rm c2}\sin{\frac{\theta}{2}}$ and the critical current is given by
\begin{eqnarray}\label{eq:Isw}
I_{\rm sw} = (2I_{\rm c1}  \zeta  + I_{\rm c2}   ) \sqrt{1- \zeta^2},
\end{eqnarray}
with $\zeta =\sqrt{I_{\rm c2}^2/8I_{\rm c1}^2 + {1}/{2}} - I_{\rm c2}/8I_{\rm c1}$ (see Appendix A for details).
On the other hand in the current decreasing stage, because the two-level system has finite probabilities on both levels due to the LZ transitions, the Josephson current changes to $I_{\rm J}=I_{\rm c1}\sin{\theta}+I_{\rm c2}(|\psi_1|^2 - |\psi_0|^2)\sin{\frac{\theta}{2}}$. For this case, the critical current would be smaller, since the two levels with opposite parities carry opposite currents and cancel each other. If the cancellation is perfect with $|\psi_1| = |\psi_0|$, the corresponding critical current is
\begin{equation}\label{eq:Ire}
I_{\rm re} = I_{\rm c1}.
\end{equation}
Consequently, a hysteresis phenomenon emerges due to the existence of the Majorana modes. We note that this hysteresis requires neither local nor global parity conservation and is immune to various quasiparticle poisoning effects in realistic setups\cite{yangpeng,lossprb,lossprb2} (see Appendix D, E for details).

The hysteresis is solely due to the nonlinear dynamics of the two-level system formed by Majorana modes. Therefore we would expect it to disappear after the topological phase transition into the trivial superconducting phase. When the system approaches the transition point from the topological nontrivial side, the spatial spreading of Majorana modes increases, which gradually annihilates the hysteresis with two mechanisms. First,
the overlapping of two Majorana modes on the same side of the junction increases, which greatly enlarges the coupling energy $\delta$. The eigenstates become states with approximately equal weight of $|0\rangle$ and $|1\rangle$. Therefore the state of this two-level system for the current-increasing and -decreasing process become approximately the same, so the hysteresis gradually disappears.
Second, the weight of the wave function of Majorana mode at the edges become smaller. Correspondingly the tunneling current of the Majorana channel $I_{\rm c2}$ decreases and so does the hysteresis.

From the classical model described by Eq. (\ref{eq:classical}), the hysteresis is similar to the mechanical hysteresis from the dry friction \cite{wojewoda} since Eq. (\ref{eq:classical}a) is actually a friction equation. That is, the particle has different friction forces when it is static and moving in the direction of $\theta$. This difference comes from the history dependent trajectories\cite{tinkhambook,Peotta14, Shev16, Guarcello17} in the $s-\phi$ plane (See Appendix B for details), and then feedback to the motion in the $\theta$ direction through the last term in Eq. (\ref{eq:classical})a. This feedback effectively induces a difference in the static friction and dynamic friction for the particle; therefore the particle would begin and stop moving at different dragging forces.

\section{Direct comparison with experiments}
In recent experiments, hysteretic I-V curves have been reported in a number of overdamped topological Josephson junctions, which are unexpected from the conventional shunted junction theory\cite{tinkhambook,molenkamp,molenkamp2}.
We argue that these hysteresis behaviors possibly come from Majorana modes as we demonstrated from the QRSJ model. In order to prove our argument, we quantitatively compare our theoretical results with experimental results.
For this reason, we consider the resistively and capacitively shunted junction model,
\begin{equation}\label{eq:RCSJ}
 I = \frac{ \hbar C{\rm d}^2 \theta} {2e{\rm d}t^2} + \frac{\hbar{\rm d} \theta}{2eR{\rm d}t}  + I_{\rm c1}\sin{\theta} + I_{\rm c2} 
 \langle i\gamma_{\rm L} \gamma_{\rm R}\rangle \sin{\frac{\theta}{2}} ,
\end{equation}
and the master equation for the two-level system\cite{wangpra}(see Appendix E for details),
\begin{equation}\label{eq:master}
\frac{{\rm d}\rho}{{\rm d}t}=-\frac{i}{\hbar}[H,\rho]+ \frac{1}{\tau_{2}}L_{2},
\end{equation}
where $\rho$ is the density matrix of the two-level system, $\tau_2$ is the decoherence time, and $L_{2} = |\psi_{{\rm g}}\rangle\langle\psi_{{\rm e}}|$ is the standard Lindblad form
where $|\psi_{{\rm e}}\rangle$ and $|\psi_{{\rm g}}\rangle$ are
the two instantaneous eigenstates of the two-level system. This combination of Eq. (\ref{eq:RCSJ}) and (\ref{eq:master}) can describe the small but nonzero capacitance and the decoherence in experiments, however, it is too complicate for analytical solution. Here we numerically simulate the model where the junction parameters are taken from the experimental data\cite{molenkamp}, with resistance $R=44\Omega$ and capacitance $C=34$aF. The Josephson current components $I_{\rm c1}=2\mu$A and $I_{\rm c2}=2.3\mu$A are extracted from the switching and retrapping current of the experimental I-V curve\cite{molenkamp}. The decoherence time is taken as $\tau_2 = 10^5\hbar/E_{\rm M}$. It is much larger than other time scales ($\tau_s,\tau_\theta,\tau_\phi$), which is reasonable because the decoherence is suppressed by the superconducting gap\cite{lossprb,lossprb2}.
The result of the simulation is presented in
 direct comparison with the experimental data as shown in Fig. \ref{fig:qrcsj}d.
Our theoretical results agree well with the experimental data.
 As far as we know, the experimental results have no convincing explanation so far, and it has never been associated with the topological nature of the junction.
Our results give a reasonable explanation for the experimentally reported ``unexpected'' hysteresis from the aspect of Majorana modes.

\begin{figure}[t]
\begin{center}
\includegraphics[clip = true, width =\columnwidth]{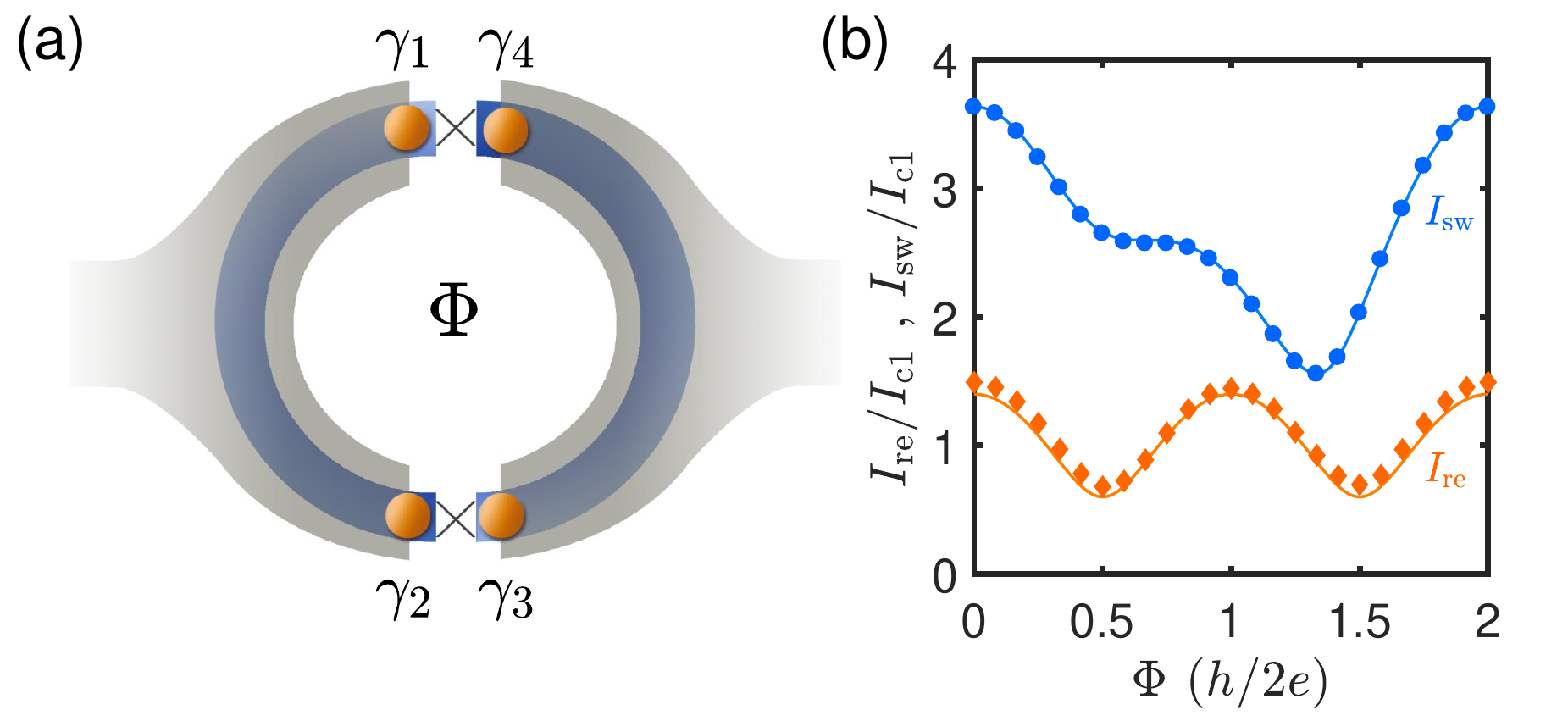}
\caption{(Color online) (a) Schematic setup of a topological SQUID structure with four Majorana zero modes. (b) The analytical interference pattern for the switching current (blue solid line) and the retrapping current (orange solid line), and the numerically results for the interference patterns of switching current (blue circle) and retrapping current (orange diamond). Josephson currents are taken as $I'_{\rm c1}/I_{\rm c1}= I'_{\rm c2}/I_{\rm c2} = 0.4$. Other Parameters are taken the same as Fig. \ref{fig:qrcsj}c for two identical junctions.}
\label{fig:squid}
\end{center}
\end{figure}

\section{Interference Pattern of a Topological SQUID.}
Hysteresis is also expected in a SQUID composed by two such junctions as shown in Fig. \ref{fig:squid}a, where the flux dependence of critical currents is a routine measurement\cite{kouwenhoven15}.
The same as for the single topological junction, the I-V curve of this SQUID should also be hysteretic. Then we expect two interference patterns of maximum supercurrent, one for the switching current and the other for the retrapping current. The switching current should contain contributions from both the conventional and Majorana channel and is thus given by
\begin{eqnarray}
I_{\rm sw} (\Phi) = \max_\theta  \big[I_{\rm c1} \sin \theta &+& I'_{\rm c1} \sin (\theta + \frac{2\pi \Phi}  {\Phi_0} ) \\\nonumber
                      + I_{\rm c2}  \sin \frac{\theta}{2} &+& I'_{\rm c2} \sin (\frac{\theta}{2}+\frac{\pi \Phi}{\Phi_0})\big],
\end{eqnarray}
where $I_{\rm c1}$ and $I_{\rm c2}$ represent the supercurrent for the quasiparticle and Majorana channels in one junction, $I'_{\rm c1}$ and $I'_{\rm c2}$ represent the supercurrent for the quasiparticle and Majorana channels in the other junction, $\Phi$ is the magnetic flux through the SQUID, and $\Phi_0 = h/2e$ is the superconducting flux quantum. Here we require the total parity conservation of the coupled Majorana modes.
This interference pattern, as shown explicitly in Fig.~\ref{fig:squid}b, is obviously $2\Phi_0$-periodic, which agrees with previous studies \cite{beenakker11,veldhorst12}.
On the other hand, the currents from Majorana channels are almost canceled when considering the retrapping current, which leads to
\begin{eqnarray}
I_{\rm re} (\Phi) \approx \max_\theta\big[I_{\rm c1} \sin \theta + I'_{\rm c1} \sin (\theta + 2\pi \Phi / \Phi_0) ],
\end{eqnarray}
which is $\Phi_0$-periodic as shown in Fig. \ref{fig:squid}b.
$I_{\rm sw}$ and $I_{\rm re}$ can be directly obtained by numerically studying the dynamics with the QRSJ model, where the Hamiltonian for the coupled Majorana modes in the SQUID is
\begin{eqnarray}
H&=& - i\gamma_1 \gamma_4 E_{\rm u} \cos({\theta}/{2})- i\gamma_2 \gamma_3 E_{\rm d} \cos[{(\theta+ 2\pi\Phi/\Phi_0)}/{2}] \nonumber\\
&&+ i\delta_{\rm l} \gamma_1 \gamma_2 + i\delta_{\rm r} \gamma_3 \gamma_4,
\end{eqnarray}
with $E_{\rm u,d}$ and $\delta_{\rm l,r}$ the corresponding coupling coefficients. The numerical results are shown in Fig. \ref{fig:squid}b, which agree well with our analytical results.

From both the analytical and numerical results, in a topological SQUID we can obtain coexistence of $h/e$ and $h/2e$-periodic interference patterns, which as far as we know is never seen in any SQUID before. The physical reason behind this phenomenon is that the Majorana channel contributes
only to switching current but negligibly to retrapping current. This unique interference phenomenon, if experimentally verified, will be an evidence for the existence of Majorana modes.

\section{Conclusion}
In summary, we propose that the Landau-Zener effect of the two-level system in a topological Josephson junction can lead to hysteresis in the I-V characteristics. We establish a quantum resistively shunted junction model to study the problem. We demonstrate the nonlinear quantum oscillation in the two-level system of the junction, with both numerical simulation and analytical methods, and show that the hysteretic I-V curves naturally follows from it. We compare our theoretical results with existing experimental results and find them quantitatively in agreement. We predict coexistence of $h/e$-periodic and $h/2e$-periodic interference patterns which are subjected to further experimental verifications.

\begin{acknowledgments}
The authors are grateful for Pavan Hosur, Stefan Ludwig, Chin-Sen Ting and Hongqi Xu for helpful discussions. This work was supported by Grants Nos. NKRDPC-2017YFA0206203, 2017YFA0303302, 2018YFA0305603,
the National Natural Science Foundation of China under Grants Nos. 11774435 and No. 61471401, and China Scholarship Council under Grants No. 201706385057. Zhao Huang is supported by Robert A. Welch Foundation under Grant No. E-1146. Qian Niu is supported by DOE (DE-FG03-02ER45958, Division of Materials Science and Engineering), NSF (EFMA-1641101) and Robert A. Welch Foundation (F-1255). 
\end{acknowledgments}


\appendix

\section{Josephson Hamiltonian and Josephson current}
Here we present a derivation for the $2\pi$-period Josephson current $I_{\rm c1}$, the $4\pi$-period Josephson current $I_{\rm c2}$, and the Hamiltonian of the two-level system $\mathcal{H}_{\rm M}$. In realistic topological Josephson junctions, usually there are both topological and non-topological segments\cite{beenakker11}. For example, in the topological superconducting nanowire as sketched in Fig. 1, the wire is topological and the substrate s-wave superconductor is non-topological. The topological segment carries the $4\pi$-period Josephson current due to Majorana modes while the non-topological segment carries the $2\pi$-period Josephson current.

Here we use a phenomenological model to describe a Josephson junction with both topological and non-topological segments. It is a hybrid two-layer system with one layer as a spinless Kitaev chain and the other layer as a trivial s-wave superconductor.

We first consider the trivial layer which is described by a simple Hamiltonian as
\begin{eqnarray}
\mathcal{H}_\alpha&&=
- t_\alpha  \sum_{ \langle i , j \rangle, \alpha, \sigma} c_{i , \alpha, \sigma}^\dagger c_{j , \alpha,\sigma}
- \mu_\alpha \sum_{ i ,  \alpha,\sigma} c_{i,  \alpha,\sigma}^\dagger c_{i, \alpha,\sigma}
\nonumber\\
&&
 +\sum_{i, \alpha} (\Delta_\alpha e^{i\theta_\alpha}  c_{i, \alpha,\uparrow}^\dag c_{i, \alpha,\downarrow}^\dag + h.c.),
\end{eqnarray}
where $\alpha={\rm L,R}$ represents the left and right sides
of the wire, $c_{\alpha,j,\sigma}$ is the electron annihilation operator
on the site $j$ and spin $\sigma = \uparrow, \downarrow$, $\Delta_{\alpha}$ is the superconductor gap, $\theta_{\alpha}$
is the superconducting phase, $t_{\alpha}$ is the nearest neighbor
hopping, and $\mu_{\alpha}$ is the chemical potential. Here for simplicity
we take identical parameters for the left and right segments, except
for the superconducting phase $\theta_{\alpha}$ which must be different
in the presence of a Josephson current.
The two superconductors are connected with a tunneling Hamiltonian,
\begin{eqnarray}
\mathcal{H}_T =  \sum_{\sigma}   (Tc_{L ,\sigma}^{\dagger} c_{R,\sigma} +h.c.) ,
\end{eqnarray}
where $c_{L ,\sigma}^{\dagger}$ is the electron creation operator at the boundary of the left superconductor nearby the junction, $T$ is the tunneling strength which is determined by the tunneling barrier of the junction. In a realistic junction, this can be controlled by an applied gate voltage.
The Josephson current can be calculated with the standard Green function technique, where the current is expressed as,
\begin{eqnarray}
I &&= 4 e T^2 {\rm Im } [  \sum_{k,p,i\omega} \Im^\dagger(k,i\omega) \Im(p, i\omega)]
\nonumber\\
&&=I_{\rm c1}\sin\theta,
\end{eqnarray}
where $\theta = \theta_{\rm L} - \theta_{\rm R}$ is the Josephson phase, $\Im$ is the off-diagonal Matsubara Green function, and $I_{\rm c1}$ is given by the contour integral as,
\begin{eqnarray}\label{eq:2picritical}
I_{\rm c1}
\approx \frac{  e  \Delta T^2}{  2 (1 - \mu^2/ 4 t^2) \hbar t^2}.
\end{eqnarray}
We note that $I_{c1}$ is a square function of $T$ which reflects the Cooper-pair tunneling. Higher order contributions in the S-matrix expansion can also be included, however, they should be negligible for the tunneling regime where the tunneling $T$ is small compared with the hopping $t$.

Now we consider topological layer, which can be studied with a spinless
p-wave superconducting Hamiltonian proposed by Kitaev\cite{kitaevaip},
\begin{eqnarray}
\mathcal{H}_{\alpha}&=&\sum_{j=1}^{N_{\alpha}}\left[-t_{\alpha}c_{\alpha,j}^{\dagger}c_{\alpha,j+1}+\Delta_{\alpha}e^{i\theta_{\alpha}}c_{\alpha,j}c_{\alpha,j+1}+h.c.\right] \nonumber \\
&&-\mu_{\alpha}\sum_{j=1}^{N_{\alpha}}c_{\alpha,j}^{\dagger}c_{\alpha,j}.
\end{eqnarray}
In this model,
the electron operators can be transformed to Majorana operators
$\gamma_{\alpha,j,{\rm A}}=e^{i{\theta_{\alpha}}/{2}}c_{\alpha,j}+e^{-i{\theta_{\alpha}}/{2}}c_{\alpha,j}^{\dagger}$
and $\gamma_{\alpha,j,{\rm B}}=-ie^{i{\theta_{\alpha}}/{2}}c_{\alpha,j}+ie^{-i{\theta_{\alpha}}/{2}}c_{\alpha,j}^{\dagger}$.
Then
the Hamiltonian can be rewritten in this Majorana representation,
\begin{eqnarray}
\mathcal{H}_{\alpha}&=&\frac{(t+\Delta)}{2}\sum_{j=1}^{N-1}i\gamma_{\alpha,j,{\rm B}}\gamma_{\alpha,j+1,{\rm A}}-\frac{(t-\Delta)}{2}\sum_{j=1}^{N-1}i\gamma_{\alpha,j,{\rm A}}\gamma_{\alpha,j+1,{\rm B}} \nonumber\\
&&-\frac{{\mu}_{\alpha}}{2}\sum_{j=1}^{N}i\gamma_{\alpha,j,{\rm A}}\gamma_{\alpha,j,{\rm B}}.
\end{eqnarray}
It is well known that this Kitaev model enters the topological non-trivial
phase for the parameter regime of $|t|>|\mu|$ and $\Delta\neq0$,
while the Majorana modes $\gamma_{{\rm L}}$, $\gamma'_{{\rm L}}$,
$\gamma_{{\rm R}}$, and $\gamma'_{{\rm R}}$ appears at the ends
of the two segments\cite{kitaevaip}. Then the low energy
(below superconducting energy gap $\Delta$) physics of the two segments
is described by an effective Hamiltonian,
\begin{eqnarray}
\mathcal{H}_{{\rm \delta}}=\sum_{\alpha}i\delta_{\alpha}\gamma'_{\alpha}\gamma_{\alpha},\label{eq:delta}
\end{eqnarray}
where $\delta_{\alpha}$ represents the coupling energy within the
left/right segment, which is exponentially protected by the length
of the wire\cite{marcus16}.

The two segments are coupled by the electron tunneling through the
barrier, which could be described by a standard tunneling Hamiltonian
\begin{eqnarray}
\mathcal{H}_{{\rm T}}=Tc_{{\rm L,N}}^{\dagger}c_{{\rm R,1}}+T^{*}c_{{\rm R,1}}^{\dagger}c_{{\rm L,N}}.
\end{eqnarray}
For low energy physics, the effective
Hamiltonian should only involve the four Majorana modes. Therefore the tunneling
Hamiltonian should be projected to these four Majorana modes with a form of\cite{kitaevaip},
\begin{eqnarray}
\mathcal{H}_{{\rm M}} &  & =-iE_{{\rm M}}\gamma_{{\rm L}}\gamma_{{\rm R}}\cos({\theta}/{2}),\label{eq:tunneling}
\end{eqnarray}
with $E_{{\rm M}}\approx T/4$ the Josephson energy. The combination of Eqs. (\ref{eq:delta})
and (\ref{eq:tunneling}) give the low energy effective Hamiltonian
of the Majorana modes in the Josephson junction, which provides a typical two-level
system. Let us look at it in more detail by defining the fermionic
operators $f_{1}=(\gamma_{{\rm L}}+i\gamma_{{\rm R}})/2$ and $f_{2}=(\gamma'_{{\rm R}}+i\gamma'_{{\rm L}})/2$
with the four Majorana modes. Then the low energy Hamiltonian can be transformed
back to the fermionic representation as,
\begin{eqnarray}
\mathcal{H} &=&\mathcal{H}_{{\rm M}}+\mathcal{H}_{\delta}\nonumber \\
&=&-E_{{\rm M}}\cos(\theta/2)(f_{1}^{\dagger}f_{1}-f_{1}f_{1}^{\dagger})\nonumber \\
&&+\delta_{{\rm L}}(f_{2}-f_{2}^{\dagger})(f_{1}+f_{1}^{\dagger})+\delta_{{\rm R}}(f_{2}+f_{2}^{\dagger})(f_{1}-f_{1}^{\dagger}).\nonumber \\
\end{eqnarray}
There are natural basis states for this Hamiltonian: $|00\rangle$,
$f_{1}^{\dagger}f_{2}^{\dagger}|00\rangle$, $f_{2}^{\dagger}|00\rangle$,
and $f_{1}^{\dagger}|00\rangle$, with $|00\rangle$ the vacuum state
for $f_{1}^{\dagger}$ and $f_{2}^{\dagger}$. With these basis states,
the total Hamiltonian can be rewritten in the matrix form as,
\begin{eqnarray}
\mathcal{H}=\left(\begin{smallmatrix}E_{{\rm M}}\cos(\theta/2) & \delta_{{\rm L}}+\delta_{{\rm R}} & 0 & 0\\
\delta_{{\rm L}}+\delta_{{\rm R}} & -E_{{\rm M}}\cos(\theta/2) & 0 & 0\\
0 & 0 & E_{{\rm M}}\cos(\theta/2) & -\delta_{{\rm L}}+\delta_{{\rm R}}\\
0 & 0 & -\delta_{{\rm L}}+\delta_{{\rm R}} & -E_{{\rm M}}\cos(\theta/2)
\end{smallmatrix}\right). \nonumber \\
\end{eqnarray}
This is a block diagonal matrix, with the left-up and right-down blocks
corresponding to the even and odd total parities, respectively. Without
losing generality, we take the even total parity and arrive at the
matrix shown in Eq. (1) of the main text with $\delta=\delta_{\rm L}+\delta_{\rm R}$.

Now let us consider the Josephson current through the Majorana
channel. The electron number operator on the right-hand side of the
junction is $N_{{\rm R}}=\sum_{j}c_{{\rm R},j}^{\dagger}c_{{\rm R},j}$,
and its time derivative gives the tunneling current,
\begin{eqnarray}
I(t)&=&-e\langle\frac{{\rm d}N_{{\rm R}}}{{\rm d}t}\rangle \nonumber\\
&=&-e\langle\psi(t)|\frac{i}{\hbar}[H,N_{{\rm R}}]|\psi(t)\rangle \nonumber\\
&=&\frac{ie}{\hbar}\langle\psi(t)|-Tc_{{\rm L,N}}^{\dagger}c_{{\rm R,1}}+T^{*}c_{{\rm R,1}}^{\dagger}c_{{\rm L,N}})|\psi(t)\rangle,\label{current}
\end{eqnarray}
where $|\psi(t)\rangle$ is the ground state wave function after including
the tunneling Hamiltonian. The single
electron tunneling through Majorana modes is obtained by the
zero-order degenerate perturbation as,
\begin{eqnarray}
I=I_{{\rm c2}}\sin({\theta}/{2})\langle\psi(t)|i\gamma_{{\rm L}}\gamma_{{\rm R}}|\psi(t)\rangle,
\end{eqnarray}
with the maximum value
\begin{eqnarray}
I_{{\rm c2}}\approx \frac{eE_{{\rm M}}}{\hbar} = \frac{eT}{4\hbar}.
\end{eqnarray}
Here we notice that $I_{{\rm c2}}$ is linear in $T$ which reflects the phase coherent single electron tunneling. Comparing with Eq. (\ref{eq:2picritical}), we obtain the ratio between the amplitude of the $2\pi$-period supercurrent and the $4\pi$-period supercurrent 
\begin{eqnarray}
\frac{I_{{\rm c1}}}{I_{{\rm c2}}} \approx \frac{  2  \Delta T}{   (1 - \mu^2/ 4 t^2)  t^2}.
\end{eqnarray}
We notice that it is a linear function of the tunneling strength $T$. That is, the $4\pi$-period supercurrent contributed by Majorana modes dominants the transport for junctions with high tunneling barriers, while the $2\pi$-period supercurrent contributed by quasiparticles dominants the transport for junctions with high transparency.

Finally we give a derivation for the switch current shown in Eq. (\ref{eq:Isw}). It is the maximum current when $s=1$ where $\theta$ is a free variable. We first calculate the Josephson phase for achieving the maximum current, which is denoted as $\theta_c$. It is obtained by taking phase derivative of the Josephson current
\begin{eqnarray}
\frac{\rm d}{\rm d\theta_c}\left(I_{\rm c1}\sin\theta_c+I_{\rm c2}\sin\frac{\theta_c}{2}\right)=0,
\end{eqnarray}
which gives
\begin{eqnarray}
\zeta \equiv \cos\frac{\theta_c}{2}= \sqrt{I_r^2+1/2}- I_r,
\end{eqnarray}
with $I_r = I_{\rm c1} / 8I_{\rm c_2}$.
We plug it back to the expression for the Josephson current and obtain
\begin{eqnarray}
I_{\rm sw} &&= I_{\rm c1}\sin\theta_c+I_{\rm c2}\sin\frac{\theta_c}{2}
\\\nonumber
 &&=\sqrt{1-\zeta^2}\left(2I_{\rm c1}\zeta+I_{\rm c2}\right),
\end{eqnarray}
which gives the Eq. (\ref{eq:Isw}).

\section{Casting the two-level system to a classical Hamiltonian}
Now we demonstrate how to cast the Schr\"{o}dinger equation for the two-level
system
\begin{eqnarray}\label{eq:SE2}
i \hbar \frac{\rm d}{{\rm d}t} \left(\begin{array}{cc}
\psi_0  \\
\psi_1
\end{array}\right) = \left(\begin{array}{cc}
E_{\rm M}\cos{\frac{\theta}{2}} & \delta\\
\delta & -E_{\rm M}\cos{\frac{\theta}{2}}
\end{array}\right) \left(\begin{array}{cc}
\psi_0  \\
\psi_1
\end{array}\right)
\end{eqnarray}
into classical equations, and form a classical dynamical
system by combining with the equation for Josephson phase from resistively shunted junction model. The wave function of the
two-level system is $(\psi_{0},\psi_{1})^{\rm T}\equiv(|\psi_{0}|e^{i\phi_{0}},|\psi_{1}|e^{i\phi_{1}})^{\rm T}$
which contains two complex numbers. However, it
obeys two constraints. First, it must be normalized $|\psi_{0}|^{2}+|\psi_{1}|^{2}=1$;
second, the overall phase of the wave function is decoupled from the dynamics of the two-level system.
With these constraints, the wave function can actually be described
by two real dynamical variables. One convenient choice 
is the relative amplitude $s\equiv|\psi_{1}|^{2}-|\psi_{0}|^{2}$
and the relative phase $\phi=\phi_{1}-\phi_{0}$. Now we derive
the equations for these two real variables out of the Schr\"{o}dinger equation.
For this purpose, we explicitly write down the amplitude and the phase of
the wave function using $s$ and $\phi$. The amplitude
of the wave function is determined by $s$ with $|\psi_{0}|=\sqrt{(1-s)/2}$
and $|\psi_{1}|=\sqrt{(1+s)/2}$, while the phase of the wave function
is determined by the relative phase $\phi$ and the total phase $\phi_{{\rm T}}=\phi_{1}+\phi_{0}$
with $\phi_{0}=(\phi_{{\rm T}}-\phi)/2$ and $\phi_{1}=(\phi_{\rm T}+\phi)/2$.
Then we can transform the Schr\"{o}dinger equation into the form,
\begin{eqnarray}
&&i \hbar \frac{\rm d}{{\rm d}t} \left(\begin{smallmatrix}
 \sqrt{\frac{1- s}{2}} e^{-i\phi/2} \\
 \sqrt{\frac{1+ s}{2}} e^{i\phi/2}
\end{smallmatrix}\right) e^{i\phi_{\rm T}/2} \nonumber\\
&=& \frac{1}{2}\left(\begin{matrix}
{E_{\rm M}}\cos{\frac{\theta}{2}} & {\delta}\\
 {\delta}& -{E_{\rm M}}\cos{\frac{\theta}{2}}
\end{matrix}\right) \left(\begin{smallmatrix}
 \sqrt{\frac{1- s}{2}} e^{-i\phi/2} \\
 \sqrt{\frac{1+ s}{2}} e^{i\phi/2}
\end{smallmatrix}\right)e^{i\phi_{\rm T}/2},
\end{eqnarray}
We note that we have added a factor of $1/2$ in front of the Hamiltonian to simplify the formula in the following derivation. Therefore both $\delta$ and $E_m$ are rescaled to be doubling their original value. We
reach at two complex equations for the real variables $s$, $\phi$, and
$\phi_{\rm T}$. The first equation is,
\begin{eqnarray}
&&i\hbar(-\sqrt{\frac{1}{8(1-s)}}\dot{s}-\frac{i}{2}\sqrt{\frac{1-s}{2}}\dot{\phi}+\frac{i}{2}\sqrt{\frac{1-s}{2}}\dot{\phi}_{\rm T}) \nonumber\\
&=&\frac{E_{{\rm M}}}{2}\cos{\frac{\theta}{2}}\sqrt{\frac{1-s}{2}}+\frac{\delta}{2}\sqrt{\frac{1+s}{2}}e^{i\phi}.
\end{eqnarray}
The imaginary part of the equation gives,
\begin{eqnarray}
\dot{s}=-\frac{\delta}{\hbar}\sqrt{1-s^{2}}\sin{\phi},\label{eq:7b}
\end{eqnarray}
which is the Eq. (\ref{eq:classical}b), while the real part of
the equation gives,
\begin{eqnarray}
\dot{\phi}-\dot{\phi}_{{\rm T}}=\frac{E_{{\rm M}}}{\hbar}\cos\theta/2+\frac{\delta \sqrt{1+s}}{\hbar \sqrt{1-s}}\cos\phi.\label{eq:7c1}
\end{eqnarray}
Checking the second equation we would have,
\begin{eqnarray}
\dot{\phi}+\dot{\phi}_{{\rm T}}=\frac{E_{{\rm M}}}{\hbar}\cos\theta/2-\frac{\delta \sqrt{1-s}}{\hbar \sqrt{1+s}}\cos\phi.\label{eq:7c2}
\end{eqnarray}
Combining the Eqs. (\ref{eq:7c1}) and (\ref{eq:7c2}), we obtain
the Eq. (\ref{eq:classical}c),
\begin{eqnarray}
\dot{\phi}=\frac{E_{{\rm M}}}{\hbar}\cos\theta/2+\frac{\delta s}{\hbar \sqrt{1-s^{2}}}\cos\phi.\label{eq:7c}
\end{eqnarray}
Rearranging the formulas we arrive at Eq. (\ref{eq:classical}). We have two equations for the two-level system,
\begin{eqnarray}\label{eqs}
\frac{{\rm d}s(t)}{{\rm d}t}&=&
 -\frac{\delta}{\hbar} \sqrt{1-s^2(t)} \sin \phi(t) \nonumber\\
&=&-\frac{1}{\tau_s} \sqrt{1-s^2(t)} \sin \phi(t) ,
\end{eqnarray}
and
\begin{eqnarray}\label{eqphi}
\frac{{\rm d}\phi(t)}{{\rm d}t}&=&\frac{E_{\rm M}}{\hbar}\cos \frac{\theta(t)}{2} + \frac{s(t) \delta }{\hbar \sqrt{1-s^2(t)}}  \cos \phi(t) \nonumber\\
&=&\frac{1}{\tau_\phi}\cos \frac{\theta(t)}{2} + \frac{s(t)  \cos \phi(t) }{\tau_s \sqrt{1-s^2(t)}} ,
\end{eqnarray}
and one equation for the Josephson phase,
\begin{eqnarray}\label{eqtheta}
\frac{{\rm d}\theta(t)}{{\rm d}t}&=&\frac{2eR}{\hbar} \left[I- {I_{\rm c1}}\sin{\theta(t)} - I_{\rm c2} s(t)\sin{\frac{\theta(t)}{2}}\right]  \nonumber\\
&=&\frac{1}{\tau_\theta} \left[1- I_1 \sin{\theta(t)} -  I_2 s(t) \sin{\frac{\theta(t)}{2}}\right],
\end{eqnarray}
where $\tau_s = \hbar/\delta$, $\tau_\phi = \hbar/E_{\rm M}$, $\tau_\theta = \hbar/2eRI$, and
we redefine two dimensionless parameters $I_1 \equiv I_{c1} / I$ and $I_2 \equiv  I_{c2}/I$ for mathematical simplicity. We see that $\phi_T$ is decoupled from these three equations.

\begin{figure}[h]
\begin{center}
\includegraphics[clip = true, width = \columnwidth]{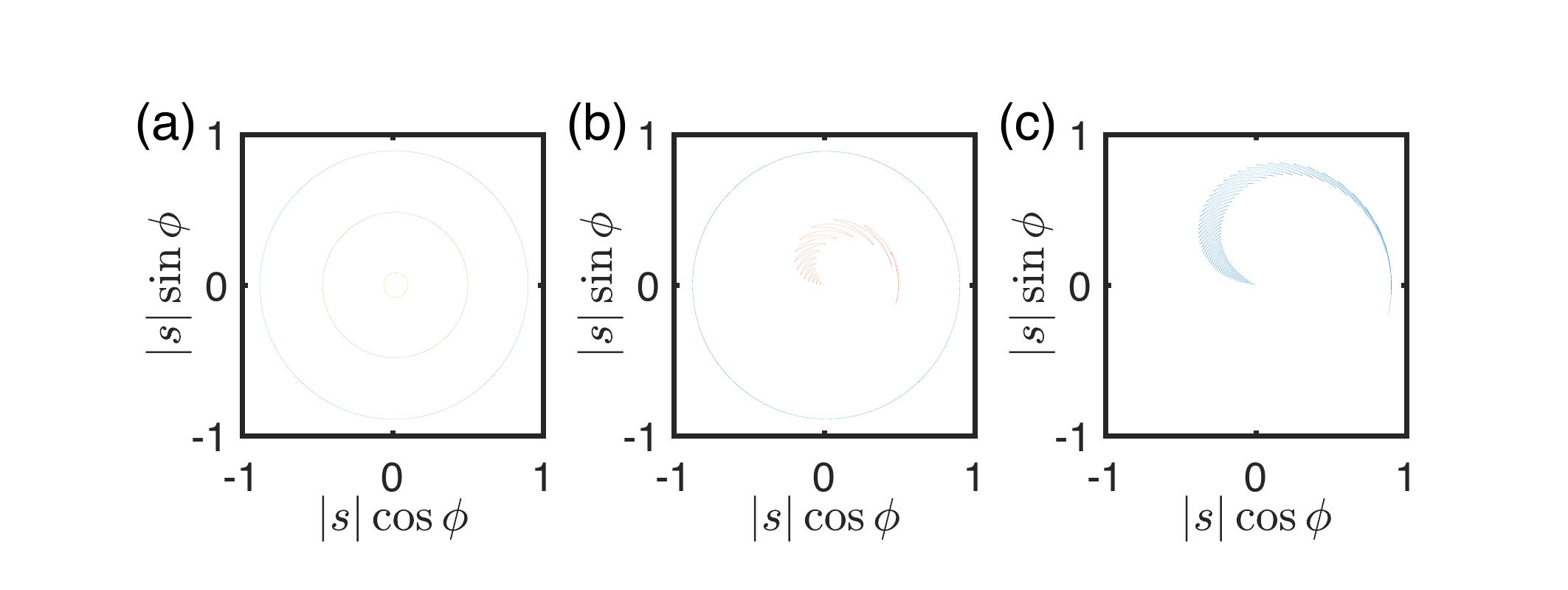}
\caption{(Color online) Typical trajectories of the particle in the phase space for the injected current of (a) $I/I_{\rm c1} = 0.5$ below the retrapping current, (b) $I/I_{\rm c1} = 2.2 $ between the retrapping current and the switching current, and (c) $I/I_{\rm c1} = 4$ above the switching current. Other parameters are taken the same as in Fig. \ref{fig:LZ}a. }
\label{fig:classical}
\end{center}
\end{figure}

Within this pure classical model, we first analyze the dynamical stability of the junction with the injected current $I$ as the control parameter. As shown in Fig. \ref{fig:classical}, we numerically explore three different injected currents. For a small current $I< I_{\rm c1}$, the trajectories for all initial conditions are closed, demonstrating circles in the $s-\phi$ plane as seen in Fig. \ref{fig:classical}a. For an intermediate current $I=2.2I_{\rm c1}$, there are two different types of trajectories, depending on the initial conditions. The trajectory for large initial $s$ is closed while the trajectory for small initial $s$ is not closed, falling to $s \approx 0$ instead. For a large current $I= 4I_{\rm c1}$, all trajectories are falling to $s \approx 0$. These results demonstrate that the dynamics for a regime of injected current depends on the initial value of $s$. This history dependence suggests the effect of nonlinearity in the dynamical evolution \cite{tinkhambook,Peotta14, Shev16, Guarcello17} and the falling of $|s|$ indicates existence of damping mechanism. In the following, we adopt the method of averaging to analytically study the Eq. (\ref{eq:classical}).

\section{Method of averaging}
After casting the QRSJ model into a purely classical model, we obtain a set of classical nonlinear equations. At first sight, the new classical equations for $s$ and $\phi$ are no simpler than the Schr\"{o}dinger equation in the original QRSJ model. However, the advantage of this pure classical formalism is the availability of sophisticated mathematical approaches that have been developed to study nonlinear classical dynamics.
From Eq. (\ref{eq:classical}) we have extracted three typical time scales
which are different by orders with $\tau_\theta \ll \tau_\phi \ll \tau_s$.
For nonlinear dynamical systems with multiple time scales, the method of averaging is a powerful mathematical tool\cite{sanders,smith}.
It was initially developed by Krylov and Bogoliubov to tackle nonlinear oscillation problems such as the study of the Einstein equation for Mercury\cite{KB}, and from then on the method has been found useful in many physical systems involving oscillations\cite{smith,sanders}.
The essence of the method of averaging is to categorize the dynamical variables as `fast' variables and `slow' variables depending on their typical time scales of variation. Then the slow variables are regarded as almost unchanged within the time scale of the fast variables, and the time dependence of the fast variables can be solved with the slow variables as fixed parameters. After obtaining this time dependence, the fast variables are averaged over time and the averaged values are plugged back into the dynamical equations for the slow variables. Finally, the time dependence of the slow variables can be solved with these averaged values of fast variables as external parameters.

The method of averaging allows us to study the dynamics of fast variables and slow variables one by one, which is much easier than investigating the full complicate coupled nonlinear equations.
In the following analysis, we can treat $\theta$ as the fast variable and $(s,\phi)$ as the slow variables. We will see that within $\tau_\theta$, the zeroth-order averaging which uses a time-independent $s$ to replace the function $s(t)$, is enough to give the high-frequency oscillation shown in Fig. 2a. The first-order averaging, where a time-independent $\dot s$ is also taken into account within $\tau_\theta$, is capable of reproducing the damping feature.

\subsection{Time Averaging over Fast Variable}
As seen in Eq. (\ref{eqphi}), the fast variable $\theta$ enters the dynamics of the slow variables through the function $\cos \theta/2$. Now, we try to calculate the time average for this function.
The whole time of dynamics can be cut into fractions of the time scale for the fast variable $\tau_\theta$. The slow variable $s(t)$ should be almost unchanged within each fraction $\tau_\theta$. As a zeroth-order averaging, $s(t)$ is treated as time independent in the equation for $\theta$. Therefore Eq. (\ref{eqtheta}) becomes,
\begin{equation}\label{eq:zeroorder}
\frac{{\rm d}\theta (t)}{{\rm d}t}= \frac{1}{\tau_\theta} \left[1 - I_1 \sin{\theta (t)} -  I_2 s  \sin{\frac{\theta (t)}{2}}\right],
\end{equation}
which can be solved alone without considering the Eqs. (\ref{eqs}) and (\ref{eqphi}) at the moment. The time evolution of $\theta$ can be obtained by solving only one differential equation, and afterwards we can make time averaging over the $\cos \theta/2$ which is defined by
\begin{eqnarray}
\overline {\cos}\frac{\theta}{2}  \equiv \frac{1}{T_\theta}\int_{0}^{T_\theta}{\rm d}t\cos\frac{\theta (t)}{2},
\end{eqnarray}
where $T_\theta$ is the time for $\theta$ to increase $4\pi$ which is at the order of $\tau_\theta$. Obviously this is a function of the parameter $s$. Here we take a simple approach to evaluate the average without solving Eq. (\ref{eq:zeroorder}) explicitly. We replace the integration over time with an integration over phase $\theta$,
\begin{eqnarray}\label{eq:zeroorderaverage}
\overline \cos\frac{\theta}{2}=
\frac{\int_{0}^{4\pi} \frac{{\rm d}\theta}{\dot \theta} \cos\frac{\theta}{2} }{\int_{0}^{4\pi}\frac{{\rm d}\theta}{\dot \theta}}=
\frac{\int_{0}^{4\pi}{\rm d}\theta\frac{\cos\frac{\theta}{2}}{1- I_1 \sin{\theta}- I_2  s \sin{\frac{\theta}{2}}}}{\int_{0}^{4\pi}{\rm d}\theta\frac{1}{1- I_1 \sin{\theta}- I_2 s \sin{\frac{\theta}{2}}}}
,
\end{eqnarray}
where the solution of $\theta(t)$ from Eq. (\ref{eq:zeroorder}) is used implicitly to accomplish the transformation. We note that the time average $\overline\cos\frac{\theta}{2}$ is nonzero because $\theta$ is not linear in time.

This expression for the time average $\overline \cos\frac{\theta}{2}$ only contains $s$, therefore corresponds to the zeroth-order averaging. Now we go to first-order averaging by including the influence of $\dot s$.
Let us derive this term by taking time derivative to Eq. (\ref{eqtheta}),
\begin{eqnarray}
\frac{{\rm d}^2\theta(t)}{{\rm d}t^2}&=&\frac{1}{\tau_\theta^2}(- I_1 \cos{\theta(t)} -  \frac{1}{2} I_2 s(t) \cos{\frac{\theta(t)}{2}})  \nonumber\\
&&*(1- I_1 \sin{\theta(t)} - I_2  s(t)  \sin{\frac{\theta(t)}{2}})   \nonumber\\
&&-\frac{1}{\tau_\theta}I_2 \dot s(t)  \sin{\frac{\theta(t)}{2}}.
\end{eqnarray}
Within $\tau_\theta$, since $s(t)$ and $\dot s(t)$ vary slowly, we consider both of them as time independent and mathematically replace the dynamical variables with static parameters $s(t) \approx s $ and $\dot s (t) \approx \dot s $, which is the first order approximation as $\dot s$ is now taken into account. We note that the time dependent velocity $\dot s (t)$ reverses the sign under time reversal operation $t \rightarrow -t$, while the parameter $\dot s$ stays the same. We thus have a plus/minus ambiguity in the replacement $\dot s (t) \approx \pm \dot s$. Now we arrive at a dynamical equation for $\theta$ as,
\begin{eqnarray}\label{eq:secondorder}
 \tau^2_\theta \frac{{\rm d}^2\theta (t)}{{\rm d}t^2} &=& \left(- I_1 \cos{\theta (t)} -  \frac{1}{2} I_2 s \cos{\frac{\theta (t)}{2}}\right)  \nonumber\\
&&*\left(1- I_1 \sin{\theta (t)}- I_2  s  \sin{\frac{\theta (t)}{2}}\right)  \nonumber\\
&&\pm  I_2 \tau_\theta \dot s  \sin{\frac{\theta (t)}{2}} \nonumber\\
&&=- \frac{\partial V (s, \dot s, \theta) } {\partial \theta },
\end{eqnarray}
with the potential function
\begin{eqnarray}
V (s, \dot s, \theta) = - \frac{1}{2} \left(1- I_1 \sin{\theta} \pm  I_2  s  \sin{\frac{\theta}{2}}\right)^2 \pm 2I_2 \tau_\theta \dot s  \cos{\frac{\theta}{2} } . \nonumber\\
\end{eqnarray}
This resembles a Newtonian equation for a particle with mass $ \tau^2_\theta $ moving under a potential $V$, therefore obeys a conservation law within each $\tau_\theta$,
\begin{eqnarray}
E = \frac{1}{2}  \tau^2_\theta \dot \theta ^2 + V (s, \dot s, \theta),
\end{eqnarray}
which gives a solution for $\dot \theta$ as
\begin{eqnarray}\label{eq:dotsaverage}
\dot{\theta}&=&\pm \frac{1}{\tau_\theta} \sqrt{2\left[E-V (s, \dot s, \theta)\right]} \nonumber\\
&=&\pm \frac{1}{\tau_\theta} \sqrt{ 2E + \left(1- I_1 \sin{\theta} - I_2  s  \sin{\frac{\theta}{2}}\right)^2 \pm 4I_2  \tau_\theta \dot s  \cos{\frac{\theta}{2} }}. \nonumber\\
\end{eqnarray}
This formula is the first-order averaging, and should recover the formula for zero-order approximation (Eq. (\ref{eq:zeroorder})) when $\dot s = 0$. This constraint requires the plus sign in front of the square root at the right hand side of Eq. (\ref{eq:dotsaverage}) and the energy to be $E=0$, which leads to
\begin{eqnarray}
\dot{\theta} = \frac{1}{\tau_\theta} \sqrt{  \left(1- I_1 \sin{\theta} - I_2  s  \sin{\frac{\theta}{2}}\right)^2 \pm 4I_2  \tau_\theta \dot s  \cos{\frac{\theta}{2} }}. \nonumber\\
\end{eqnarray}
With this formula for $\dot \theta$, we can analytically calculate the time average by transforming the integration over time to the integration over $\theta$,
\begin{eqnarray} \label{eq:firstorderaverage}
\overline \cos\frac{\theta}{2}&=&\frac{\int_{0}^{4\pi} \frac{{\rm d}\theta}{\dot \theta} \cos\frac{\theta}{2} }{\int_{0}^{4\pi}\frac{{\rm d}\theta}{\dot \theta}} \nonumber\\
&=&\frac{\int_{0}^{4\pi}{\rm d}\theta\frac{\cos\frac{\theta}{2}}{\sqrt{ \left(1- I_1 \sin{\theta} - I_2  s  \sin{\frac{\theta}{2}}\right)^2 \pm 4I_2  \tau _\theta \dot s  \cos{\frac{\theta}{2} }}}}{\int_{0}^{4\pi}{\rm d}\theta\frac{1}{\sqrt{ \left(1- I_1 \sin{\theta} - I_2  s  \sin{\frac{\theta}{2}}\right)^2 \pm 4I_2  \tau_\theta \dot s  \cos{\frac{\theta}{2} }}}}.
\end{eqnarray}

These two integral expressions Eq. (\ref{eq:zeroorderaverage}) and Eq. (\ref{eq:firstorderaverage}) give the time average over the fast variable up to zeroth-order and first-order averaging. Certainly we can go further to include the influence of $\ddot s$, {\it etc}. However, we find that the $s$ and $\dot s$ dependence is enough to qualitatively understand the dynamics. We will use these two integral expressions to obtain an explicit function and plug it back into the dynamical equations for the slow variables.

\subsection{The Zeroth-order Averaging and the Classical Hamiltonian}
Let us first examine the zeroth-order averaging where the time average is given by Eq. (\ref{eq:zeroorderaverage}).
Now we calculate the integrals by taking Taylor expansions,
\begin{eqnarray}
&&\frac {1}{1- I_1 \sin\theta- I_2 s \sin\frac{\theta}{2}} \nonumber\\
&=&1+ ( I_1 \sin \theta+ I_2 s \sin\frac{ \theta}{2})+(I_1 \sin \theta+ I_2 s \sin\frac{ \theta}{2})^{2} \nonumber\\
&&+(I_1\sin \theta+ I_2 s \sin\frac{ \theta}{2})^{3}+...,
\end{eqnarray}
which gives the lowest order result for the denominator of Eq. (\ref{eq:zeroorderaverage}),
\begin{eqnarray}
\int_{0}^{4\pi}{\rm d}\theta\frac{1}{1- I_1 \sin\theta- I_2 s \sin\frac{\theta}{2}}
   \approx \int_{0}^{4\pi}{\rm d}\theta  =  4 \pi.
\end{eqnarray}
Similarly we have the numerator of Eq. (\ref{eq:zeroorderaverage}) as,
\begin{eqnarray}
&&\int_{0}^{4\pi}{\rm d}\theta\frac{\cos\frac{\theta}{2}}{1-I_1\sin\theta-I_2 s \sin\frac{\theta}{2}}  \nonumber\\
&=&\int_{0}^{4\pi}{\rm d}\theta\cos\frac{\theta}{2}\left[1+(I_1\sin\theta+I_2 s \sin\frac{\theta}{2})\right.  \nonumber\\
&&\left.+(I_1\sin\theta+I_2 s \sin\frac{\theta}{2})^{2}+...\right].
  \nonumber\\
&\approx& 2 I_1 I_2 s  \int_{0}^{4\pi} d\theta  \cos\frac{\theta}{2}  \sin \frac{\theta}{2} \sin \theta  \nonumber\\
&=&  2 \pi I_1 I_2 s.
\end{eqnarray}
For both integrations, we take the lowest order nonzero term in the expansion series.
Putting the results for numerator and denominator together, we obtain
\begin{eqnarray}
\overline \cos \frac{\theta}  {2} \approx \frac{I_1 I_2}{2} s  \equiv \alpha s
\end{eqnarray}
where for simplicity we define a parameter $\alpha=I_{1}I_{2}/2$.

Now we come to the essence of the method of averaging. We replace $\cos\frac{\theta}{2}$ in Eq. (\ref{eqphi}) with the time averaged function $\overline \cos\frac{\theta}{2}$, and obtain the equations soly for $s$ and $\phi$ as
\begin{eqnarray}
\frac{{\rm d}s}{{\rm d}t}&&=-\frac{1}{\tau_s} \sqrt{1-s^2} \sin \phi,
\\\nonumber
\frac{{\rm d}\phi}{{\rm d}t}&&= \frac{\alpha s}{\tau_\phi}+ \frac{s \cos \phi }{\tau_s \sqrt{1-s^2}}  .
\end{eqnarray}
With $\theta$ averaged out, obviously these two equations are self consistent equations. In fact they are the canonical equations of a classical Hamiltonian,
\begin{eqnarray}
H_{\rm c}&=&- \frac{1 }{ 2 \tau_\phi}\alpha  s^2 + \frac{1}{\tau_s} \sqrt{1-s^2} \cos \phi,
\end{eqnarray}
where $s$ and $\phi$ are the extended coordinate and the canonical momentum. This classical Hamiltonian is the Eq. (\ref{eq:classicalH}) which represents a classical integrable system, with the evolution of the phase-space motions of the Hamiltonian shown in Fig. \ref{fig:LZ}b.

Let us examine basic features of this classical Hamiltonian. We first study the fixed points $(s_c,\phi_c)$, which are obtained by taking the stationary condition of the Hamilton equations,
\begin{eqnarray}
 \frac{{\rm d}s}{{\rm d}t} \bigg|_{s_{\rm c},\phi_{\rm c}} = -\frac{1}{\tau_s}\sqrt{1-s_{\rm c}^2} \sin \phi_{\rm c} = 0
\\\nonumber
 \frac{{\rm d}\phi}{{\rm d}t} \bigg|_{s_{\rm c},\phi_{\rm c}} =\frac{1}{\tau_\phi} {E_{\rm M}\alpha s_{\rm c}} + \frac{s_{\rm c}\cos \phi_c}{ \tau_s \sqrt{1-s_{\rm c}^2}} = 0.
\end{eqnarray}
Considering the fact that $\delta < E_{\rm M}\alpha$, there are three sets of fixed points,
\begin{eqnarray}
(s_{\rm c},\phi_{\rm c})=
\left\{
             \begin{array}{lr}
             (0,  \quad 0),  \\
             (0,  \quad \pm \pi), & \\
             (\pm \sqrt{1 - {\tau_\phi^2}/({\tau_s \alpha)^2}} ,  \quad \pm \pi).&
             \end{array}
\right.
\end{eqnarray}
The first two sets of fixed points are marked as $P_1$, $P_2$ in the \ref{fig:LZ}b. Then
we check the classification of these fixed points, which is described by the Jacobian matrix at the fixed points,
\begin{eqnarray}
J (s_{\rm c},\phi_{\rm c})&=&
\left(
\begin{array}{lr}
             \frac{\partial \dot s}{\partial s} &   \frac{\partial \dot s}{\partial \phi}\\
             \frac{\partial \dot \phi}{\partial s} &   \frac{\partial \dot \phi}{\partial \phi}
             \end{array}
\right)_{s= s_{\rm c},\phi = \phi_{\rm c}} \nonumber\\
&=&
\left(
\begin{array}{lr}
             \frac{ s_c \sin \phi_c}{\tau_s \sqrt{1-s_c^2}} &   -\frac{1}{\tau_s}\sqrt{1-s^2_c} \cos \phi_c\\
             \frac{ \alpha}{\tau_\phi} + \frac{ \cos \phi_c}{\tau_s \sqrt{1-s_c^2}} + \frac{ s^2 \cos \phi_c }{\tau_s ({1-s_c^2})^{3/2}} &   -\frac{ s_c \sin \phi_c}{\tau_s \sqrt{1-s_c^2}}
             \end{array}
\right). \nonumber\\
\end{eqnarray}
For the fixed point $P_1$ at the position $(s_c,\phi_c) = (0,0)$, we have the Jacobian matrix of
\begin{eqnarray}
J(0,0) = \left(
\begin{array}{lr}
             0&   -\frac{1}{ \tau_s}\\
            \frac{\alpha }{\tau_\phi}+ \frac{1}{\tau_s}  &  0
             \end{array}
\right),
\end{eqnarray}
This stability matrix has two imaginary eigenvalues $\lambda_{1,2} = \pm {i}\sqrt{(\alpha\tau_s + \tau\phi)/\tau_s^2\tau_\phi}$, signifying that $P_1$ is an elliptic fixed point.  Similarly, we calculate eigenvalues of the Jacobian matrices at the other two fixed points, and find that the $P_2$ is a hyperbolic fixed point, while $(s_{\rm c},\phi_{\rm c}) = (\pm \sqrt{1 - {\tau_\phi^2}/({\tau_s \alpha)^2}}, \pm \pi)$ are elliptic fixed points. These analytical results agree with the information we see on the phase space portrait shown in Fig. \ref{fig:LZ}b.

In classical dynamics, the phase-space trajectory which connects the hyperbolic fixed points is called separatrix.
In this classical Hamiltonian Eq. (\ref{eq:classicalH}), a separatrix connects the fixed points $P_2$, as shown in Fig. \ref{fig:LZ}b. The separatrix divides the phase space into distinct regions, where the evolution of the phase-space motion orbits around different elliptic fixed points. 
In the action-angle formalize\cite{dittrich},
the phase space area enclosed by an orbit defines the action,
\begin{eqnarray}
I (H_{\rm c}) = \frac{1}{2\pi} \int s(H_{\rm c},\phi){\rm d}\phi,
\end{eqnarray}
where the $s$ is a function of $\phi$ and the energy $H_{\rm c}$ by reversing Eq. (\ref{eq:classicalH}). This action is an adiabatic invariant and its derivative on energy gives the period of oscillation
\begin{eqnarray}
T = \frac{{\rm d}I (H_{\rm c})}{{\rm d}H_{\rm c}}.
\end{eqnarray}
Since the hyperbolic fixed points locate at energy saddle points, the separatrix has the diverging energy derivative. Therefore, we would expect a slow down of the oscillation if the motion is going near to the separatrix, which is clearly seen in Fig. \ref{fig:LZ}a.

The classical Hamiltonian Eq. (\ref{eq:classicalH}) obtained by the method of averaging captures a number of features of the simulation results. The phase space of the classical Hamiltonian is divided into two distinct areas by the separatrix. The orbits outside the separatrix only have small oscillations in $s$ with its value remains positive or negative, while the orbits inside the separatrix can oscillate between negative minimums and positive maximums. These two distinct types of orbits agree with the dynamics of the $s$ shown in Fig. \ref{fig:LZ}a, with the oscillation first small and only at the negative value, and later becoming large and between negative and positive values. Interestingly, right at the transition between these two distinct oscillating behaviors, we observe obvious enlargement of the period in Fig. \ref{fig:LZ}a. This indicates that the system is walking through the separatrix which has the divergent period. Comparing the Fig. \ref{fig:LZ}a and Fig. \ref{fig:LZ}b, it is reasonable to argue that the oscillation behaviors are well described by the effective classical Hamiltonian, but the damping of the oscillating amplitude cannot be understood yet.

One direct method to view the resemblance between the classical Hamiltonian and the original system is to draw the Poincar\'e map for the evolution of the motion obtained by numerically solving the QRSJ model. The Poincar\'e map is the intersections of a chosen surface in the phase space, called as the Poincar\'e surface of section, and the motion trajectories in the whole phase space\cite{dittrich}. This approach replaces the integration of equations with the study of mappings, and has shown much power in nonlinear dynamics. It is particularly advantageous in understanding the qualitative features of the system. In our present case, we naturally choose the $s-\phi$ plane with $\theta = 0$ as the Poincar\'e surface of section since we hope to compare it with the two-dimensional phase space portrait of the effective classical Hamiltonian. The obtained Poincar\'e map is shown in Fig. \ref{fig:LZ}c. We see that the structure of the
Poincar\'e map resembles the phase-space portrait of the effective Hamiltonian. In particular, we see circling features in the Poincar\'e map, which looks similar to the elliptic orbits around the fixed point $P_1$ for the classical Hamiltonian. However, if examined more carefully, the points actually spiral to the fixed point $P_1$. This is consistent with the large number of points around $P_1$, which indicate the convergence of the trajectories and the breaking down of the phase-space volume conservation.

\begin{figure}[h]
\centering{}\includegraphics[clip,width=\columnwidth]{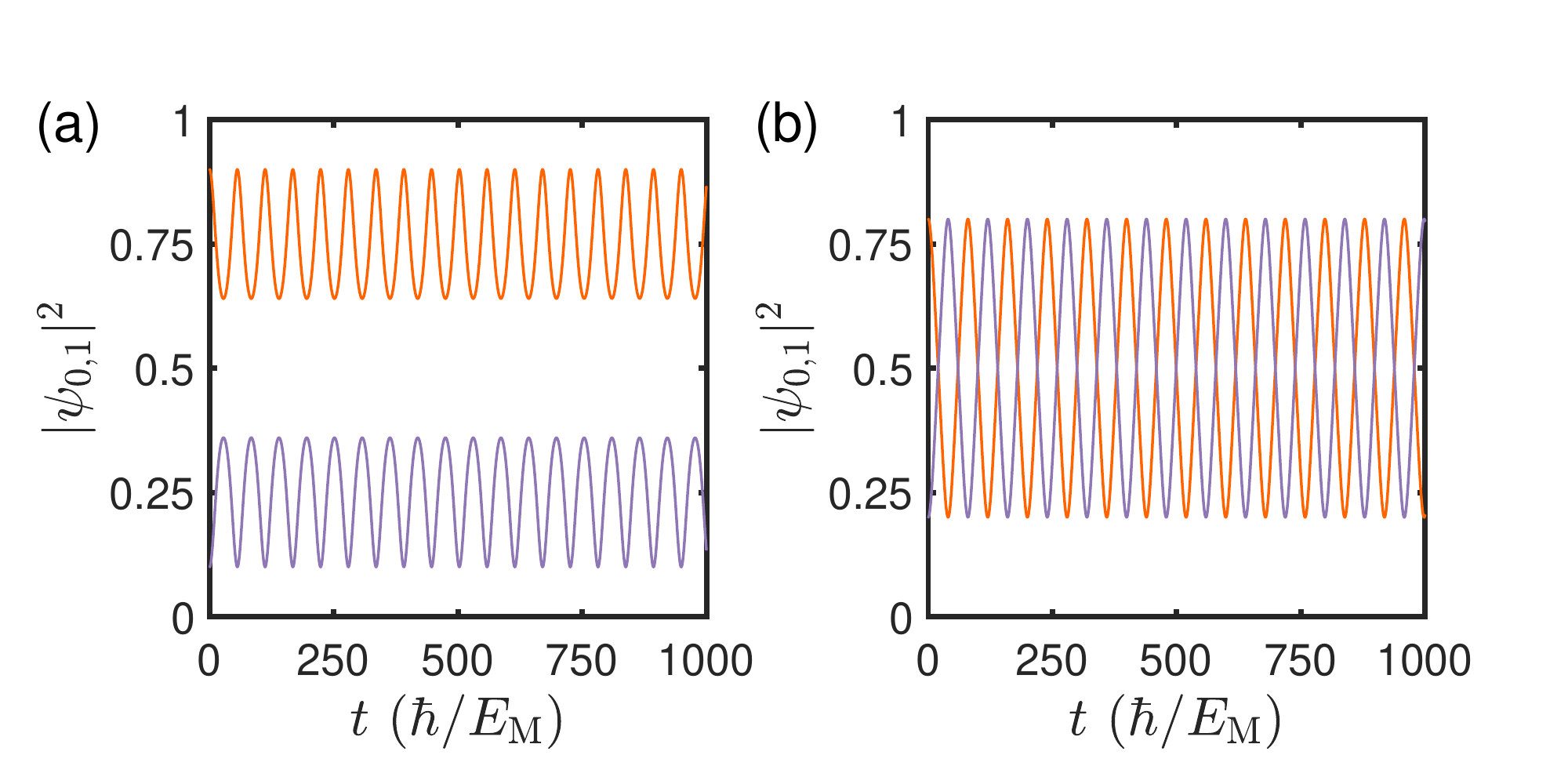}
\caption{(Color online) The time evolution of the wave function simulated for the nonlinear Schr\"{o}dinger equation Eq. (\ref{eq:nonlinearSE}), with initial condition of (a) $\psi_0 = \sqrt{0.9}, \psi_1 = \sqrt{0.1}$, and (b) $\psi_0 = \sqrt{0.8}, \psi_1 = \sqrt{0.2}$. Other parameters are taken the same as Fig. \ref{fig:LZ}.}
\label{fig:oscillation}
\end{figure}

Finally, it is inspiring to directly looking at the Schr\"{o}dinger equation by replacing $\cos\frac{\theta}{2}$ with the averaging $\overline \cos\frac{\theta}{2}$ in the Eq. \ref{eq:SE}. We can obtain the Eq. (\ref{eq:nonlinearSE}), which is a nonlinear Schr\"{o}dinger equation where other nontrivial LZ phenomena have been discussed before\cite{Liujie,liujie2}. From this equation, we see that the coupling between the two-level system and the Josephson phase naturally induces the nonlinearity to the quantum dynamics, which is the origin of such a rich and unusual dynamics for the wave function of the two-level system as shown in Fig. \ref{fig:LZ}a. The dynamical of this nonlinear Schr\"{o}dinger equation is initial value dependent, as expected from the equivalent classical Hamiltonian.
We numerically simulate the time evolution of the wave function with two typical initial values, and show the results in Fig. \ref{fig:oscillation}. We find that the wave function exhibits oscillation patterns similar to the oscillations at different time ranges in Fig. \ref{fig:LZ}a. However, the damping of the oscillating amplitude is missing, which will be explained in the following section.

\subsection{The first-order averaging and the damped harmonic oscillator}
The classical Hamiltonian Eq. (\ref{eq:classicalH}) helps us to understand the quantum oscillation of the two-level system. However, it can not describe the damping of the oscillation as shown in Fig.\ref{fig:LZ}a. From the Poincar\'e map shown in Fig. \ref{fig:LZ}c, we know that the damping is towards the elliptic fixed point $P_1$. The natural guess is that it comes from an extra friction force which is proportional to the velocity of the extended coordinate $\dot s$. This could be obtained from the expression of the first-order averaging Eq. (\ref{eq:firstorderaverage}).
Similar to the calculation of the zeroth-order averaging, the lowest-order Taylor expansion for the integrated function gives the result
\begin{eqnarray}\label{eq:dotaverageresult}
\overline \cos\frac{\theta}{2}  =  \alpha s + \beta \dot{s}  + O (s^2, \dot s^2),
\end{eqnarray}
where $\beta \dot s $ represents the small contribution from the first-order averaging with $\beta = \pm I_2 \tau_\theta $. If we plug this averaging result back to Eq. (\ref{eq:classical}c), we will find that the second term is linear in $\dot s$, thus adding a velocity dependent force beyond the classical Hamiltonian Eq. (\ref{eq:classicalH}).
For classical mechanical systems where forces only depend on coordinates, the Liouville theorem guarantees the phase-space volume conservation, which guarantees undamped oscillations. On the other hand, the existence of the velocity dependent force stemming from the $\beta \dot s$ term in Eq. (\ref{eq:dotaverageresult}) breaks the Liouville theorem and the phase-space volume conservation. This is why the Poincar\'e map in Fig. \ref{fig:LZ}c shows a phase-space volume compression, leading all trajectories toward the elliptic fixed point $P_1$.

Now we explicitly show that this $\dot s$ dependent term induces damping to the oscillation. For this purpose, we need to decouple equations for $s$ and $\phi$. Taking time derivative on both sides of Eq. (\ref{eq:classical}b), we obtain
\begin{eqnarray}
\ddot{s} = -\frac{\delta }{\hbar } \left[ - \frac{s \sin \phi} {\sqrt{1-s^2}} \dot s + \sqrt{1-s^2} \cos \phi \dot \phi
\right].
\end{eqnarray}
Then we put Eqs. (\ref{eq:classical}b) and (\ref{eq:classical}c) into the right side of this equation to eliminate the $\dot s$ and $\dot \phi$ and arrive at
\begin{eqnarray}
\ddot{s} &=&-\frac{\delta }{\hbar } \left[ - \frac{s \sin \phi} {\sqrt{1-s^2}} \left( -\frac{\delta}{\hbar} \sqrt{1-s^2} \sin \phi\right) \right.  \nonumber\\
&&\left.+ \sqrt{1-s^2} \cos \phi \left(\frac{E_{\rm M}}{\hbar}\cos \frac{\theta}{2} + \frac{s \delta }{\hbar \sqrt{1-s^2}}  \cos \phi \right)\right] \nonumber\\
&=&-\frac{\delta }{\hbar } \left[
\frac{\delta}{\hbar} s + \frac{E_{\rm M}}{\hbar}\cos \frac{\theta}{2} \sqrt{1-s^2} \cos \phi
\right].
\end{eqnarray}
To eliminate $\phi$ from the right side of the equation, we notice that Eq. (\ref{eq:classical}b) can be transformed with trigonometric identity as
\begin{eqnarray}
\dot s^2 = \frac{\delta^2}{\hbar^2} \left[(1-s^2) - (\sqrt{1-s^2} \cos \phi )^2 \right],
\end{eqnarray}
which can be used to replace the $\phi$ depend term and we obtain
\begin{eqnarray}
\ddot{s} =-\frac{ 1 }{\tau_s^2 } s
 \pm \frac {1 } {\tau_s \tau_\phi} \cos \frac{\theta} {2} \sqrt{1-s^2 - (\tau_s\dot s)^2},
\end{eqnarray}
where the ambiguity of the plus/minus sign comes from taking the square root. Now we plug the averaging result Eq. (\ref{eq:dotaverageresult}) into the equation, and obtain
\begin{eqnarray}\label{eq:sfriction}
&&\ddot{s} + \left(\frac{1}{\tau_s^2}  + \frac{1}{\tau_s \tau_\phi} \alpha\sqrt{1-s^2 - (\tau_s \dot s)^2}\right) s \nonumber\\
&\approx&-\frac{I_2\tau_\theta}{\tau_s \tau_\phi} \sqrt{1-s^2 - (\tau_s \dot s)^2}\dot s ,
\end{eqnarray}
where the correct plus/minus signs are chosen for obtaining consistent results with numerical simulations. This second order differential equation represents a damped oscillator, where the angular frequency and the damping ratio depend on $s$. The damping comes from the right side of the equation which is a friction term proportional to $\dot s$.

 \begin{figure*}[t]
\centering{}\includegraphics[clip,width=2\columnwidth]{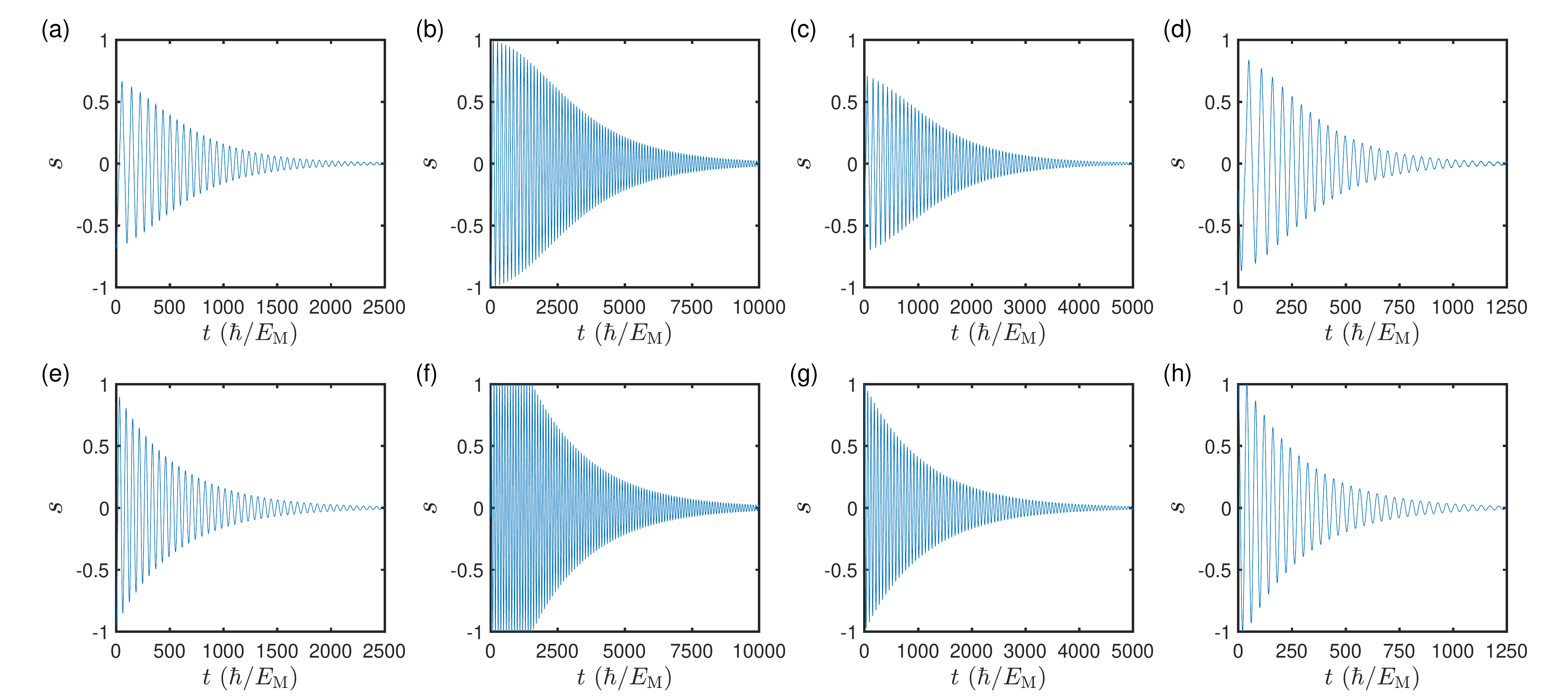}
\caption{(Color online) Comparison between numerical results and the analytical solutions shown in Eq. (\ref{eq:dampsolution}) from the equations of damped harmonic oscillators. (a) The numerical results with the same parameters as Fig. \ref{fig:LZ}a but taken from the time range of $t$ = [3830,6330] in the original figure. The origin of time is shift to zero for comparison. (b) The numerical results with the same parameters as in (a) except for $I=6I_{c1}= 3I_{c2}$, with data taken between the time range [250,10250]. (c) The numerical results with the same parameters as in (a) except for $R= 20\hbar/(2e^2)$, with data taken between the time range [7000,12000]. (d) The numerical results with the same parameters as in (a) except for $\delta= 0.04 E_{\rm M}$, with data taken between the time range [650,1900].
(e-h) The analytic solution shown in Eq. (\ref{eq:dampsolution}) with parameters taken the same as in (a-d) respectively. The time origins in each figure have been shifted accordingly to make the initial value comparable to numerical results.}
\label{fig:analyticcompare}
\end{figure*}

By considering the regime of $s, \tau_s\dot s \ll 1 $, we have the approximation $\sqrt{1-s^2 - (\tau_s \dot s)^2}\approx 1$. Then Eq. (\ref{eq:sfriction}) can be further simplified to
\begin{eqnarray}
\ddot{s} + \left(\frac{1}{\tau_s^2}  + \frac{1}{\tau_s \tau_\phi} \alpha
\right) s  + \frac{I_2\tau_\theta}{\tau_s \tau_\phi}   \dot s = 0,
\end{eqnarray}
which has exactly the same form as a classical damped harmonic oscillator. It can be rewritten to the standard form of,
\begin{eqnarray}\label{eq:dampedoscillator}
\ddot{s}+ 2\xi  \omega_0 \dot s + \omega_0^2 s= 0,
\end{eqnarray}
with an angular frequency of
\begin{eqnarray}
\omega_0^2 = \frac{1}{\tau_s^2}  + \frac{\alpha}{\tau_s \tau_\phi},
\end{eqnarray}
and a damping ratio of
\begin{eqnarray}
\xi =  \frac{I_2 \tau_\theta}{2\tau_\phi\sqrt{1+\alpha \tau_s/\tau_\phi}}.
\end{eqnarray}
This damped harmonic oscillator is underdamped with a small damping ratio $\xi \ll 1$ due to $\tau_\theta \ll \tau_\phi$. Finally we arrive at the solution for $s$ around the elliptic fixed point as,
\begin{eqnarray}\label{eq:dampsolution}
s (t) = e^{-t/\tau_{\rm d}} \cos (t/\tilde \tau_s),
\end{eqnarray}
with
\begin{eqnarray}
\tau_{\rm d}&=& \frac{1}{\xi \omega_0}= \frac{2\tau_s \tau_\phi}{ I_2 \tau_\theta}  , \nonumber\\
\quad \tilde \tau_s &=& \frac{1}{\omega_0 \sqrt{1-\xi^2}}\approx \frac{1}{\omega_0 }  =  \frac{\tau_s} {\sqrt{ 1  + \alpha {\tau_s}/{ \tau_\phi} } },
\end{eqnarray}
where we used $\xi \ll 1$ in the second formula. These two time scales characterize the slow damping and the fast oscillation of $s$. The new time scale $\tau_{\rm d}$ is the largest time scale which can be constructed from the three basic time scales of the system. We compare the analytical solution given by Eq. (\ref{eq:dampsolution}) with the numerical results directly from Eq. (\ref{eq:classical}) for several different sets of parameters, as shown in Fig. \ref{fig:analyticcompare}. We find quantitative agreement between them when $s$ approaches zero.

From above, we find that the quantum dynamics of the two-level system is dual to the classical dynamics of a damped harmonic oscillator after adopting the method of averaging. This helps us to successfully obtain an analytical solution of the quantum dynamics of the two-level system within the separatrix of the effective Hamiltonian. This dual relation gives a new insight in studying quantum two-level systems when nonlinearity is introduced.

\subsection{Anharmonic damped oscillator}
The solution of the damped harmonic oscillator obtained in Eq. (\ref{eq:dampsolution}) only becomes accurate when $s$ approaches zero. Here we show an improved approximation which works at larger $s$. Based on the numerical results and the analytical solution Eq. (\ref{eq:dampsolution}), we know that the solution is a form of damped oscillation. Therefore, we propose an ansatz solution of the form
\begin{eqnarray}
s=A (t) \cos(t/\tau_s' ),
\end{eqnarray}
where $A(t)$ is the slow damping amplitude and $\tau_s'$ is the oscillating period. With this ansatz solution, we can simplify the square root term in Eq. (\ref{eq:sfriction}) to
\begin{eqnarray}
&&\sqrt{1-s^2 - (\tau_s \dot s)^2} \nonumber\\
&=&\sqrt{1-\left[A  \cos(t/\tau_s' )\right]^2 - \tau^2_s[ \dot A  \cos(t/\tau_s') - \frac{A}{\tau_s'} \sin(t/\tau_s')]^2} \nonumber\\
&\approx&\sqrt{1-A^2},
\end{eqnarray}
where in the second line we use the fact that $\dot A / \tau_s \ll 1$ and $\tau_s / \tau_s' \sim 1$ within the separatrix. Now the Eq. (\ref{eq:sfriction}) is simplified to an anharmonic damped oscillator
\begin{eqnarray}
\ddot{s}+ \left(\frac{1}{\tau_s^2}  + \frac{1}{\tau_s \tau_\phi} \alpha
\sqrt{1-A^2}\right) s \approx - \left (\frac{I_2\tau_\theta}{\tau_s \tau_\phi} \sqrt{1-A^2} \right) \dot s. \nonumber\\
\end{eqnarray}
This equation is more precise than the simple damped harmonic oscillator Eq. (\ref{eq:dampedoscillator}) since the square root is treated with a better approximation than rudely taken as unity. Now we try to obtain the analytical solution of this damped anharmonic oscillator with appropriate approximation. We first calculate $\tau'_s$ by treating $A$ as a constant within $\tau'_s$ and ignore the friction term. These two approximations are valid because $A$ varies much slower than $\tau'_s$ and the friction is ignorable in the time scale of $\tau'_s$. Then we can obtain a harmonic oscillating equation,
\begin{eqnarray}
\ddot{s} + \left(\frac{1}{\tau_s^2}  + \frac{1}{\tau_s \tau_\phi} \alpha
\sqrt{1-A^2}\right) s \approx 0,
\end{eqnarray}
which gives the oscillating period as
\begin{eqnarray}
\tau_s' = \frac{\tau_s}{\sqrt{1 + \frac{\tau_s}{ \tau_\phi} \alpha
\sqrt{1-A^2}}}.
\end{eqnarray}
Comparing with the oscillating period $\tilde \tau_s$ obtained from the damped harmonic oscillator approximation, the new oscillating period $\tau_s'$ depends on the oscillating amplitude $A$. When $A$ increases, the oscillating period $\tau'_s$ becomes larger. This agrees with the numerical results shown in Fig. \ref{fig:LZ}a, and also agrees with the analysis based on the classical Hamiltonian which states that the oscillating frequency becomes larger when approaching the separatrix.

 \begin{figure*}[t]
\centering{}\includegraphics[clip,width=2\columnwidth]{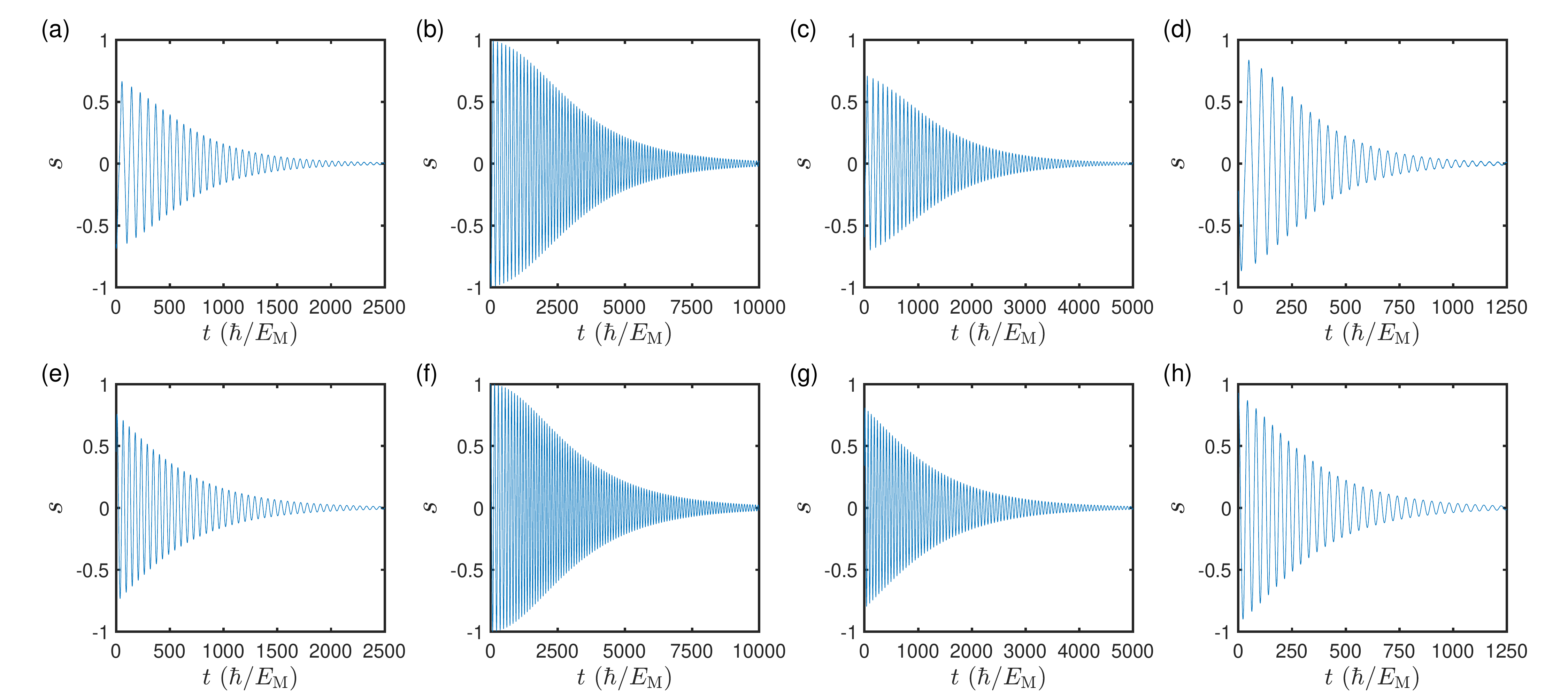}
\caption{(Color online) Comparison between numerical results and the analytical solutions shown in Eq. (\ref{eq:analyticalfinial}). (a-d) are the same as the (a-d) in Fig. \ref{fig:analyticcompare}.
(e-h) The analytic solution shown in Eq. (\ref{eq:analyticalfinial}) with parameters taken the same as in (a-d) respectively. The time origins in each figure are shifted accordingly to make the initial value comparable to numerical results.}
\label{fig:analyticcompare2}
\end{figure*}

Now we calculate the slow damping amplitude $A(t)$ by plugging the ansatz solution back to the equation,
\begin{eqnarray}
&&\ddot A \cos(t/\tau_s' ) -\frac{2 \dot A}{\tau_s'} \sin (t/\tau_s' ) \\\nonumber
&=& - \left (\frac{I_2\tau_\theta}{\tau_s \tau_\phi} \sqrt{1-A^2} \right) \left(\dot A \cos(t/\tau_s' ) - \frac{A}{\tau_s'} \sin (t/\tau_s' )\right).
\end{eqnarray}
Noticing the fact that $\ddot A(\tau_s')^2 \ll \dot A\tau_s' \ll A$ for slow varying $A$, we obtain the equation for $A$ as
\begin{eqnarray}
\frac{{\rm d}A}{{\rm d}t} &\approx& - \left (\frac{I_2\tau_\theta}{2\tau_s \tau_\phi} \sqrt{1-A^2} \right) A
\nonumber\\
&=&  - \frac{1}{\tau_{\rm d}} A \sqrt{1-A^2},
\end{eqnarray}
which has the solution
\begin{eqnarray}\label{eq:exponentialdamped}
A(t) =   \frac{ 2 e^{- t/\tau_{\rm d}}}{1+  e^{-2t/\tau_{\rm d}}}.
\end{eqnarray}
Comparing with the result of damped harmonic oscillator approximation in the previous section, this damping function is more flat when $s$ becomes large.

Putting the expression for $A(t)$ and $\tau_s'$ together, we finally arrive at the analytical solution for $s(t)$ inside the separatrix of the phase-space,
\begin{eqnarray}\label{eq:analyticalfinial}
s (t) & =&   \frac{2e^{-t/\tau_{\rm d}}}{1+e^{-2t/\tau_{\rm d}}} \cos ({t}/{\tau_s'}).
\end{eqnarray}
Clearly this solution reduces back to Eq. (\ref{eq:dampsolution}) when $s$ approaches zero. However, it provides a better result for the larger $s$ regime which captures two more details of the damped oscillation shown in Fig. \ref{fig:LZ}a. First, the oscillating period is larger when $s$ is larger, which also agrees with our analysis based on the action-angle formalism. Second, the damping of the oscillating amplitude is slower at larger $s$, which is different from the pure exponential decay which has a time-independent decay rate. We show the comparison of the numerical results with the analytical result Eq. (\ref{eq:analyticalfinial}) in Fig. \ref{fig:analyticcompare2}, and find better agreement with numerical results than the simple damped harmonic oscillator approximation.

\subsection{Krylov-Bogoliubov method of averaging}
Finally, we take an alternative method, the Krylov-Bogoliubov averaging method\cite{KB,smith}, to calculate the damping function $A(t)$, and show identical results as from the damped anharmonic oscillator approximation. We examine the Eq. (\ref{eq:sfriction}) again and make it dimensionless as,
\begin{eqnarray}\label{eq:sdynam}
\frac{{\rm d}s^{2}}{{\rm d}\tau^{2}}+s = - \frac{\tau_s}{\tau_\phi} (\alpha s +  \frac{\beta}{\tau_s}\frac{{\rm d}s}{{\rm d}\tau}) \sqrt{1-s^{2}-(\frac{{\rm d}s}{{\rm d}\tau})^{2}},
\end{eqnarray}
where we define a dimensionless time as $\tau = t / \tau_s$.
Setting the right hand side of the equation to be zero we obtain,
\begin{eqnarray}\label{eq:harmonics}
\frac{{\rm d}s^{2}}{{\rm d}\tau^{2}}+s = 0.
\end{eqnarray}
This equation has the general solution of the form
\begin{eqnarray}
s &=& A' \cos (\tau + B),
\\\nonumber
\dot s &=& - A' \sin (\tau + B ).
\end{eqnarray}
Now we recover the right hand side of the equation, and take an anartz solution with the same trigonometric functions where $A'$ and $B'$ become time dependent,
\begin{subequations}\label{eq:ansatz}
\begin{eqnarray}
s  &=& A' (\tau)\cos(\tau+B'(\tau))
\\
\dot s  &=& -A'(\tau) \sin(\tau+B'(\tau)).
\end{eqnarray}
\end{subequations}
In the following, we solve Eq. (\ref{eq:sdynam}) with these ansatz functions. We first take time derivative to Eq. (\ref{eq:ansatz}a) and obtain,
\begin{eqnarray}
\dot s = -A' \sin(\tau+B') + \dot A' \cos(\tau+B') - A'  \sin(\tau+B') \dot B' .\nonumber\\
\end{eqnarray}
This equation must be equivalent to  Eq. (\ref{eq:ansatz}b) for a self-consistent ansatz function, and thus we have a constraint equation
\begin{eqnarray}\label{eq:constraint1}
\dot A' \cos(\tau+B')=A'\sin(\tau+B') \dot B'.
\end{eqnarray}
We then plug the ansatz functions Eq. (\ref{eq:ansatz}) back to the original equation (\ref{eq:sdynam}) and obtain another constraint equation,
\begin{eqnarray}\label{eq:constraint2}
&&- \dot A' \sin(\tau+B')-A'\cos(\tau+B') \dot B' \nonumber\\
&=&\frac{E_{{\rm M}}}{\delta}(\alpha A' \cos(\tau+B') +I_2 \tau_\theta A' \sin(\tau+B')) \nonumber\\
&&*\sqrt{1-A'^{2}\cos^{2}(\tau+B')\!-\!A'^{2}\sin^{2}(\tau+B')}\nonumber\\
&=&\frac{E_{{\rm M}}A'}{\delta}(\alpha \cos(\tau+B') + I_2 \tau_\theta \sin(\tau+B'))\sqrt{1-A'^{2}}. \nonumber\\
\end{eqnarray}
Combining these two constraint equations (\ref{eq:constraint1}) and (\ref{eq:constraint2}), we arrive at equations for $A' (\tau)$ and $B' (\tau)$ as,
\begin{eqnarray}\label{eq:AB}
&&\frac{\rm d}{{\rm d}\tau}
\left(\begin{array}{cc}
A'\\
B'
\end{array}\right) \nonumber\\
 &=& -\frac{E_{{\rm M}}}{\delta}\left(\alpha \cos(\tau+B') + I_2 \tau_\theta \sin(\tau+B')\right) \nonumber\\
 &&*\sqrt{1-A'^{2}}
\left(\begin{array}{cc}
 A' \sin(\tau+B')\\
 \cos(\tau+B')
\end{array}\right).
\end{eqnarray}
We note that no approximation has been made yet. The ansatz functions Eq. (\ref{eq:ansatz}) together with the constraint equations Eq. (\ref{eq:AB}) give an {\it exact} solution
to the Eq. (\ref{eq:sdynam}).
Now we concentrate on the slow varying part of $A'$, denoting as $A$, which captures the slow damping of the oscillation. We replace $\cos(\tau+B')$ and $\sin(\tau+B')$ with their average values within one period and obtain
\begin{eqnarray}
\frac{{\rm d} A}{{\rm d}\tau} &=&- \frac{E_{{\rm M}}}{2\pi \delta} A \sqrt{1-A^{2}} \nonumber\\
&&*\int_0^{2\pi} d\tau \left[\alpha \cos(\tau+B) + \frac{I_2\tau_\theta}{\tau_s} \sin(\tau+B) \right] \sin(\tau+B) \nonumber\\
&=&   - \frac{I_2\tau_\theta  }{2 \tau_\phi } A \sqrt{1-A^{2}}.
\end{eqnarray}
Transforming back to real time with $\tau = t/\tau_s$ and rearranging the parameters, we simplify the equation of $A$ to the form
\begin{eqnarray}\label{eq:damped}
\frac{{\rm d} A}{{\rm d}t}  =  -  \frac {A}{\tau_{\rm d}} \sqrt{1-A^{2}},
\end{eqnarray}
which is exactly the same as we obtained from the damped anharmonic oscillator approximation.

\section{Hysteresis with external parity flipping}
\begin{figure}[t]
\centering{}\includegraphics[clip,width=\columnwidth]{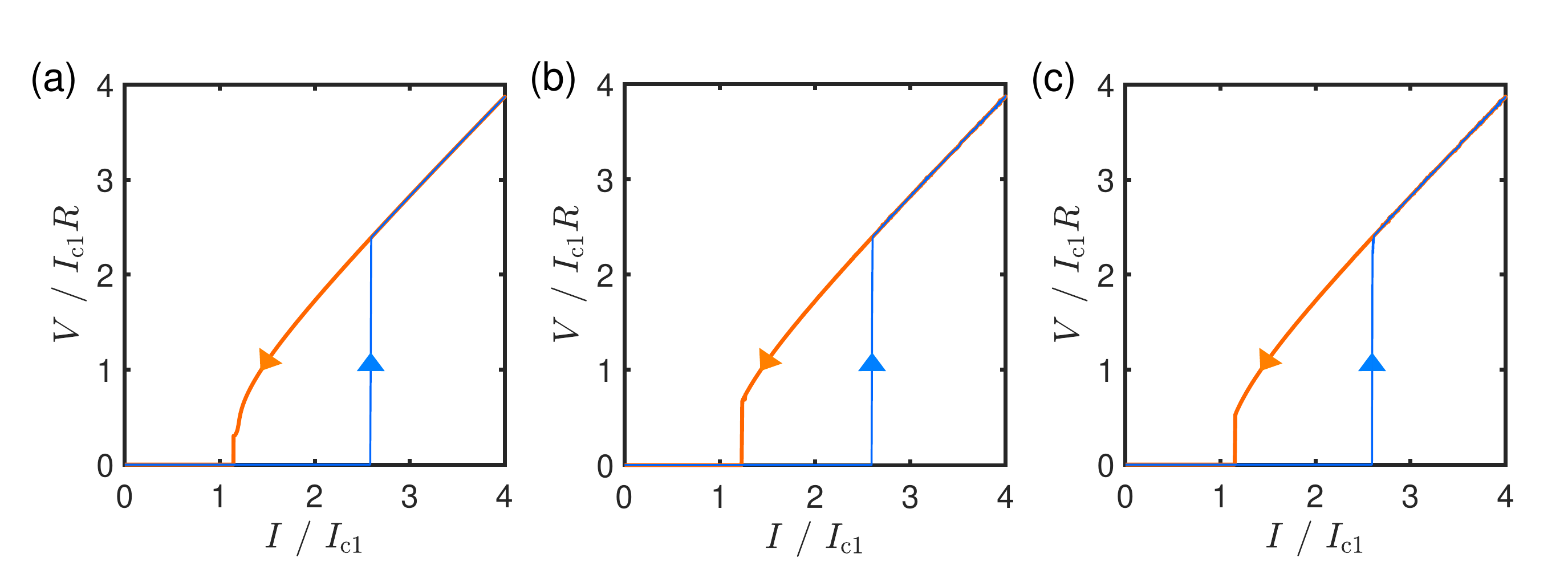}
\caption{(Color online) Numerical simulation of the I-V curves with the energies of the external quantum level as (a)$\epsilon=0$, (b)$\epsilon/E_{{\rm M}}=0.5$, (c)$\epsilon/E_{{\rm M}}=-0.5$. Parameters are taken as $T/E_{{\rm M}}=0.04$, $\delta_{\rm L}/E_{{\rm M}}=0.005$ and $\delta_{\rm R}/E_{{\rm M}}=0.015$ and other parameters are taken the same as Fig. \ref{fig:LZ}a.}
\label{fig:parityflip}
\end{figure}

Here we show that the hysteresis in the I-V curve still exits even
if the total parity of Majorana modes is broken by external quantum levels from
a single quasiparticle or impurity. For a model study, we consider
the simplest case of an extra quantum level with a Hamiltonian of
\begin{eqnarray}
\mathcal{H}_{i}=\epsilon d^{\dag}d,
\end{eqnarray}
where $\epsilon$ is the energy of the level which is near zero, and $d^{\dag}$ is the creation operator on the level.
This quantum level couples with one Majorana mode through the tunneling
Hamiltonian,
\begin{eqnarray}
\mathcal{H}_{{\rm T}} & = & T\gamma_{{\rm L}}d+T^{*}d^{\dag}\gamma_{{\rm L}}\nonumber \\
 & = & (f_{1}^{\dagger}+f)(Td-T^{*}d^{\dag}),
\end{eqnarray}
where $T$ is the tunneling strength. After including this quantum level,
the Hilbert space is expanded and the total Hamiltonian is an eight-by-eight matrix. It is also block diagonal with two four-by-four blocks due to the conservation of the total parity. We can
take one block by picking the basis states as, $d^{\dag}|00\rangle$,
$d^{\dag}f_{1}^{\dagger}f_{2}^{\dagger}|00\rangle$, $f_{2}^{\dagger}|00\rangle$,
$f_{1}^{\dagger}|00\rangle$. Then we arrive at an effective Hamiltonian
\begin{eqnarray}
\mathcal{H}=\left(\begin{smallmatrix}\epsilon+E_{{\rm M}}\cos(\theta/2) & \delta_{{\rm L}}+\delta_{{\rm R}} & 0 & T^{*}\\
\delta_{{\rm L}}+\delta_{{\rm R}} & \epsilon-E_{{\rm M}}\cos(\theta/2) & T^{*} & 0\\
0 & T & E_{{\rm M}}\cos(\theta/2) & -\delta_{{\rm L}}+\delta_{{\rm R}}\\
T & 0 & -\delta_{{\rm L}}+\delta_{{\rm R}} & -E_{{\rm M}}\cos(\theta/2)
\end{smallmatrix}\right).\label{eq:4x4} \nonumber\\
\end{eqnarray}
The quantum average for
the supercurrent through the Majorana channel is given by
\begin{eqnarray}
\langle\psi|i\gamma_{2}\gamma_{3}|\psi\rangle=|\psi_{3}(t)|^{2}-|\psi_{2}(t)|^{2}+|\psi_{1}(t)|^{2}-|\psi_{0}(t)|^{2}. \nonumber\\ \label{eq:4x4current}
\end{eqnarray}
We plug the Eqs. (\ref{eq:4x4}) and (\ref{eq:4x4current}) into the
QRSJ model, and numerically obtain the I-V curve of the junction
as demonstrated in Fig. \ref{fig:parityflip}. Clearly, the hysteresis
behavior is insensitive to the parity flipping from the external quantum
level.

The reason that the parity flipping does not change the hysteresis is that the Hamiltonians for the odd total Majorana parity (the left-up 2x2 block) and the even total Majorana parity (the right-down 2x2 block) are qualitatively similar. They both have avoided crossings at the Josephson phase $\theta = (2n+1) \pi$. Naturally, we would expect that the quantum dynamics within each block is qualitatively the same, presenting a damped oscillation. The small flipping energy $T$ will not change this quantum dynamics, therefore will not change the hysteresis behavior.

\section{Quasiparticle poisoning}
In the topological superconductors, the quasiparticle poisoning is
an important obstacle for many signatures of Majorana modes. The difference between the quasiparticle poisoning and a simple external
quantum level from impurity or quantum dot is that the quasiparticle poisoning comes from the
thermal equilibrium fermionic environment which brings decoherence
into the quantum two-level system defined by Majorana modes. This decoherence is fundamental
from the quantum mechanical point of view, and cannot be simply equivalenced
to an enlarged Hilbert space. Then it is a natural question whether
the decoherence from the quasiparticle poisoning will destroy the
LZ effect induced hysteresis. We analyze this problem by
considering the density matrix $\rho(t)=\rho_{11}(t)|0\rangle\langle0|+\rho_{12}(t)|0\rangle\langle1|+\rho_{21}(t)|1\rangle\langle0|+\rho_{22}(t)|1\rangle\langle1|$
for the two-level system where the decoherence can be naturally included using
the Lindblad form. The dynamics of the two-level system is then described by a
master equation\cite{wangpra},
\begin{equation}
\frac{{\rm d}\rho}{{\rm d}t}=-\frac{i}{\hbar}[H,\rho]+\sum_{i}\frac{1}{\tau_{i}}L_{i},
\end{equation}
where $L_{i}$ are all possible Lindblad forms which describe the
decoherence and $\tau_{i}$ are the corresponding decoherence times.
For a general two-level system, there are only three possible Lindblad forms $L_{1}=|\psi_{{\rm e}}\rangle\langle\psi_{{\rm g}}|$,
$L_{2}=|\psi_{{\rm g}}\rangle\langle\psi_{{\rm e}}|$, and $L_{3}=|\psi_{{\rm e}}\rangle\langle\psi_{{\rm e}}|-|\psi_{{\rm g}}\rangle\langle\psi_{{\rm g}}|$,
where $|\psi_{{\rm e}}\rangle$ and $|\psi_{{\rm g}}\rangle$ are
the two instantaneous eigenstates of the two-level system. When considering the
decoherence from the quasiparticle poisoning, only the relaxation
processes described by $L_{2}$ and the dephasing processes described
by $L_{3}$ are relevant in the low temperature limit.

\begin{figure}[t]
\centering{}\includegraphics[clip,width=0.9\columnwidth]{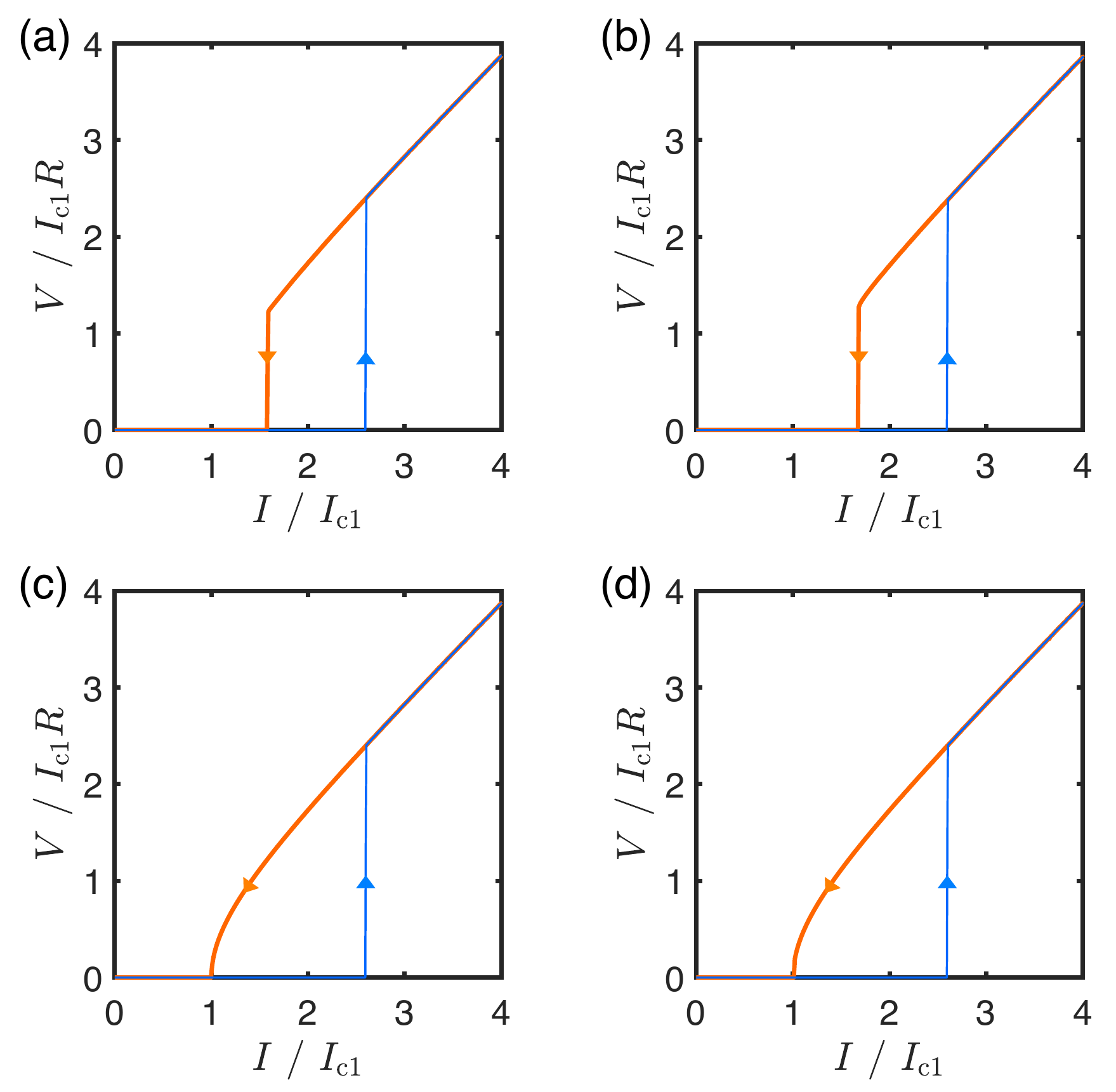} 
\caption{(Color online) Numerical results for the I-V curve with the decoherence time (a)
$\tau_{2}=1000\hbar/E_{{\rm M}}$, (b) $\tau_{2}=10\hbar/E_{{\rm M}}$,
(c) $\tau_{3}=0.1\hbar/E_{{\rm M}}$, and (d) $\tau_{2}=1000\hbar/E_{{\rm M}}$
and $\tau_{3}=0.1\hbar/E_{{\rm M}}$. Other parameters are taken the
same as Fig. \ref{fig:LZ}a.}
\label{fig:qp}
\end{figure}

Let us first consider the relaxation processes given by the Lindblad
$L_{2}$, which involves the coupling between the Majorana modes and the quasiparticle
states above the superconducting gap. The decoherence time for this
process is an exponential function of the superconducting gap\cite{lossprb,lossprb2},
\begin{equation}
\frac{1}{\tau_{2}}=\lambda Te^{-\Delta/T},
\end{equation}
where $\lambda_{0}$ is a dimensionless factor estimated around $0.01$
for quasiparticle poisoning processes in nanowire systems\cite{lossprb}. When the
temperature is far below the superconducting gap $T\ll\Delta$, the
relaxation time is exponentially protected by the superconducting
gap and would be quite long compared with all other time scales in
the system. We present the results of the I-V curve with two different
relaxation times of in Figs. \ref{fig:qp}a and \ref{fig:qp}b. We
see that a reasonable long relaxation time has little influence on the hysteresis, while an extremely short relaxation time reduces the hysteresis but still does not change
the qualitative feature.

We then consider the decoherence from the dephasing given by the Lindblad
$L_{3}$. Different from the relaxation, the dephasing should have
a relatively short dephasing time\cite{lossprb,lossprb2} with $\tau_{3}\ll\tau_{2}$.
However, looking at the form of $L_{3}$ we see that the dephasing
only introduces a decoherence in the relative phase of the two eigenstates,
leaving the relative amplitude unchanged. Since only the amplitude
of the wave function enters the dynamical equation for the Josephson phase in the QRSJ model, we
would expect that the dephasing has little influence on the hysteresis.
We present the I-V curve for a very short dephasing time in Fig. \ref{fig:qp}c,
and find that it indeed has no influence on the hysteresis behavior.
Finally, we show the result with a combination of the relaxation and
dephasing in Fig. \ref{fig:qp}d, and find that the hysteresis is
robust to the decoherence from the quasiparticle poisoning.

\section{underdamped junction}

\begin{figure}[t]
\centering{}\includegraphics[clip,width=\columnwidth]{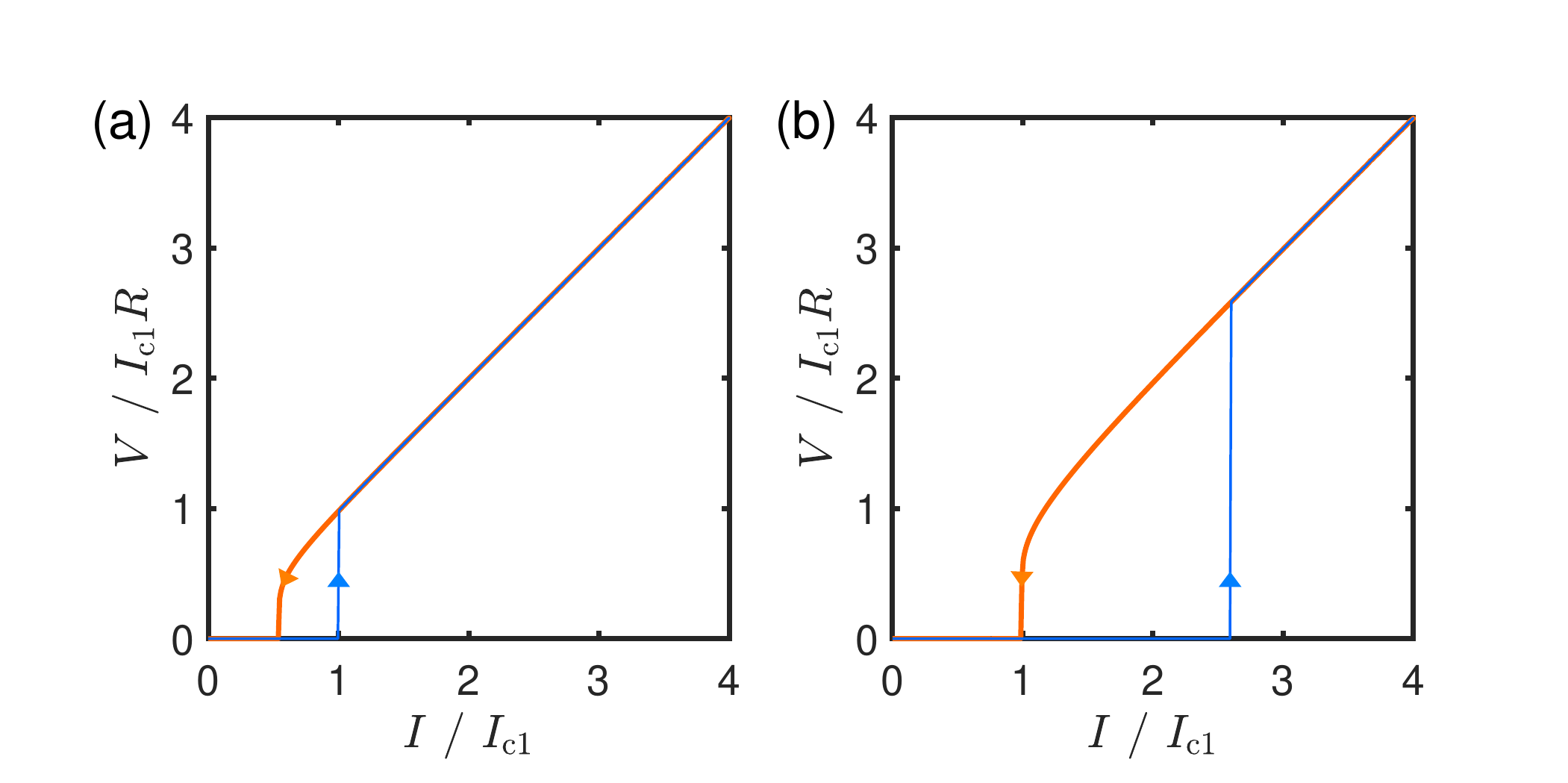}
\caption{(Color online) Numerical results of the I-V curves for the underdamped
junctions with (a) $I_{{\rm c2}}=0$ and (b) $I_{{\rm c2}}=2I_{{\rm c1}}$.
The capacitance is taken as $C=0.1e^3/\hbar I_{\rm c1}$. Other parameters are the
same as Fig. \ref{fig:LZ}a.}
\label{fig:underdamp}
\end{figure}

The conventional Josephson junctions with negligible capacitance show no hysteresis, making the
LZ effect induced hysteresis a novel phenomenon. However, even in
the underdamped junctions where hysteresis is already expected from
the shunted capacitance, the LZ effect still contribute
a significant feature which might be useful for experimental detection.
Here, we demonstrate a comparison between the I-V curves of conventional
and topological junctions in the underdamped regime, where the capacitance
is included and the resistively shunted junction equation is rewritten as the resistively and capacitively shunted junction equation.
We show the numerical results in Fig. \ref{fig:underdamp}. There
is a hysteresis in the topological trivial junction as expected from
the standard theory, however, the difference between the switching
and retrapping current is largely enhanced by the LZ effect induced
part. Therefore, it is still a useful signal for detecting the Majorana modes
in possible topological junctions.

\end{document}